\documentstyle[aps,prc,12pt,eqsecnum,tighten]{revtex}

\begin{document}

\title{\begin{flushright}
NT@UW-99-9\\
March  1999
\end{flushright}
Infinite Nuclear Matter on the Light Front: Nucleon-Nucleon
  Correlations}

\author{G.A. Miller}
\address{Department of Physics, Box 351560, University of Washington,
Seattle, Washington 98195-1560}
\author{R. Machleidt}
\address{Department of Physics, University of Idaho, Moscow, Idaho 83844}

\maketitle

\begin{abstract}

A relativistic light front formulation of nuclear
dynamics is developed and applied to treating infinite nuclear matter
in a method which includes the correlations of pairs of nucleons:
this is light front Brueckner theory. 
We start with a hadronic meson-baryon Lagrangian that
is consistent with chiral symmetry. This is used to obtain
a light front version of a one-boson-exchange nucleon-nucleon
potential (OBEP). 
 The accuracy of our description of the 
nucleon-nucleon (NN)
data is good, and similar to that of other relativistic OBEP models.
We derive, within the light front formalism, the
Hartree-Fock and Brueckner Hartree-Fock equations. 
Applying our light front OBEP, the nuclear matter
saturation properties
are reasonably well reproduced. We obtain
a value of the compressibility, 180 MeV,
that is  smaller than that of
alternative relativistic approaches to nuclear matter in which
the compressibility usually comes out too large.
Because the  derivation starts from a  meson-baryon
Lagrangian, we are able to show that replacing the meson degrees of freedom
by a NN interaction is a consistent   approximation, and the
formalism allows one to calculate corrections to this approximation in a
well-organized manner.  The simplicity of the vacuum
in our light front approach
is an important feature in allowing the derivations to proceed.
The mesonic Fock space components of the nuclear wave
function are obtained also, and aspects of the
meson and nucleon  plus-momentum distribution functions
are computed. We find that   there are about 0.05 excess pions
       per nucleon. 
\end{abstract}


\newpage

\section{Introduction}

We introduce a light front formalism for infinite nuclear matter,
in which the effects of
correlations are taken into account. This is  a light front
Brueckner theory of nuclear matter. While the 
 ultimate goal of this and related  studies
is to provide a  fully relativistic
treatment
of nuclei which includes all previous knowledge about nuclear dynamics,
the present work represents 
one step beyond the previous light-front mean-field calculation
\cite{gam97a,gam97b} of the  properties of infinite nuclear matter.


 Understanding
an important class of experiments  seems to require  that
light-front dynamics and the related  light
cone variables be used. Consider  the  lepton-nucleus deep inelastic scattering
 experiments
\cite{emc} which showed 
 that
there is a significant difference between the parton distributions
of free nucleons and nucleons in a nucleus. This difference
can interpreted as a small ($\sim$ 10\%)
shift in  the momentum distribution of
valence quarks towards smaller values of the Bjorken variable
$x_{Bj}$. The
Bjorken variable is a ratio of the plus-momentum $k^+=k^0+k^3$
of a quark to that 
of the target. If one regards
the nucleus as a collection of nucleons, $x_{Bj}
=p^+/k^+$, where  $k^+$ is the 
plus momentum of a nucleon bound in the nucleus.
 If one uses
 $k^0+k^3$ as a momentum variable the corresponding canonical
 spatial variable is $x^-=x^0-x^3$ and the time variable is $x^0
 +x^3$\cite{notation}. 
This  is  the light front (LF)
approach of Dirac\cite{Di 49}; see the recent reviews\cite{lcrevs,hari}
for more information.

Deep inelastic scattering depends on
the light-front momentum distribution which is the probability 
$f(k^+)$ that a bound nucleon has a momentum $k^+$. Other nuclear reactions,
such as (e,e') and (p,2p) depend also on this very same
probability\cite{fs1,fs2,jif}.  The quantity 
$f(k^+)$ is simply related to the square of the ground state wave
 function, computed using light front dynamics. 
The usual equal time approach
to nuclear dynamics is very successful, and it is natural to use this
information to calculate the distribution $f(k^+)$. However, in the
standard equal time
formulation this quantity is a response function and depends 
on matrix elements between the ground and
all excited states, and therefore can be more difficult to compute. 

 The use of light front
variables is convenient for interpreting certain  experiments, but
does not allow one to avoid the necessary task of handling nuclear dynamics.
Thus one is faced with the task of computing the ground state nuclear wave
function using $x^+=x^0 +x^3$ as a time variable. 
The present effort is a simplification in that the nucleus is taken to be
infinite nuclear matter. However, the detailed
effects of the interactions between two nucleons are included, so that we
are concerned with the relativistic dynamics of a strongly interacting many
body system. 

Light front techniques have previously been applied to systems of
two hadrons \cite{lcrevs,fs1,fs2,coester,Ke 91,Ka 88,Fu 91,bakker}.
Our emphasis here is in large nuclear systems.
The light front quantization procedure necessary to treat nucleon
interactions with scalar and vector mesons was derived by
Soper\cite{des71}, and by Yan and
collaborators\cite{yan12,yan34}.

We next outline our procedure. The necessary Lagrangian, which respects chiral
symmetry,  and its light front
Hamiltonian is described in Sect.~II. Its application to nucleon-nucleon
scattering in the one-boson exchange approximation is carried out in
Sect.~III. A new feature is that  the effects of isovector mesons such as
the $\bbox{\rho}$ and  $\bbox{\delta}$, and the $\bbox{\rho}$-nucleon
tensor interaction  are included. The NN potential is
generated using the one-boson exchange approximation.
The Weinberg-type integral equation, which maintains unitarity and
boost invariance in one direction, is solved and the results are compared
with phase shift data. Sect.~IV is concerned with the many-nucleon
problem. Two separate perturbation series are involved. The first step is
to eliminate temporarily
the meson degrees of freedom in favor of our two nucleon
potential. Thus one first proceeds using only nucleon degrees of freedom.
The light-front
formalism to obtain the  nucleonic interacting
ground state wave function $\mid\Phi\rangle$
in terms of a series in which a Brueckner G matrix acts on
a best Slater determinant $\mid\phi\rangle$ is developed. The independent-pair
approximation is used.
One must find an eigenstate of the $P^-$ operator for which the expectation
value of $P^+$ is equal to the eigenvalue of $P^-$.
This formalism is applied and the results are discussed in Sect.~V.
The full nuclear wave function $\mid\Psi\rangle$,  including the meson
degrees of freedom, is discussed in Sect.~VI. The object  $\mid\Psi\rangle$
is related to $\mid\Phi\rangle$ by a second  
series involving the difference between
the  nucleon-nucleon interaction and the meson nucleon interactions. One finds
that the expression for the nuclear mass, evaluated in Sect.~V, is valid within
our approximation. Furthermore, expressions for the meson and nucleon
distribution functions are obtained. We derive a sum rule
for obtaining the total number of (non-vector) mesons in the nucleus.
A brief discussion of the implications of our results for lepton-nucleus deep
inelastic scattering and the nuclear Drell-Yan process is presented in
Sect.~VII.  A  brief summary  of
our results  is contained in Set.~VIII.
Some of the necessary notation is discussed in an Appendix.
Some of the present results, but none of the details of the derivation or of
our two-nucleon potential have
appeared in Ref.\cite{mm98}. 

\section{Light Front Quantization: Lagrangian, Field equations, and Light-Front
  Hamiltonian}

The light front approach is a three-dimensional formalism involving
a Hamiltonian which is a $P^-$ operator. One starts with a Lagrangian
and derives field equations which
 allow one to eliminate the appearance of 
dependent degrees of freedom in the Hamiltonian.
Our starting point is 
 a non-linear chiral
model in which the 
nuclear constituents are nucleons $\psi$ (or $\psi')$, pions $\bbox{\pi}$,
 scalar mesons $\phi$\cite{scalar}, and
 vector mesons
$V^\mu$. 
The  Lagrangian ${\cal L}$ is given by 
\begin{eqnarray}
{\cal L} ={1\over 2} (\partial_\mu \phi \partial^\mu \phi-m_s^2\phi^2) 
-{1\over  4} V^{\mu\nu}V_{\mu\nu} +{m_v^2\over 2}V^\mu V_\mu 
+{1\over  4}f^2Tr (\partial_\mu\;U\;\partial^\mu\;
U^\dagger) \nonumber\\
+{1\over  4}m_\pi^2f^2\;
Tr(U +U^\dagger-2)
+\bar{\psi}^\prime\left(\gamma^\mu
({i\over 2}\stackrel{\leftrightarrow}{\partial}_\mu
-g_v\;V_\mu) -
U(M +g_s\phi)\right)\psi' \label{lag}
\end{eqnarray}
where the bare masses of the nucleon, scalar and vector mesons are given by 
$M, m_s,$  $m_v$, and  $V^{\mu\nu}=
\partial ^\mu V^\nu-\partial^\nu V^\mu$. The unitary  matrix $U$ can be
chosen from amongst three forms $U_i$:
\begin{equation}
U_1\equiv e^{i  \gamma_5 \bbox{\tau\cdot\pi}/f},\quad
U_2\equiv{1+i\gamma_5\bbox{\tau}\cdot\bbox{\pi}/2f\over
1-i\gamma_5\bbox{\tau}\cdot\bbox{\pi}/2f},\quad
U_3=\sqrt{1-\pi^2/f^2}+i\gamma_5\bbox{\tau\cdot\pi}/f, \label{us}
\end{equation}
which correspond to different definitions of the fields.
This Lagrangian, based on the linear representations
of chiral symmetry used by  Gursey \cite{gursey},
is discussed in Ref.~\cite{gam97b}.
It is approximately  
 $(m_\pi\ne 0$)
invariant under the chiral transformation
\begin{eqnarray}
\psi^\prime\to e^{i \gamma_5 \bbox{\tau}\cdot\bbox{a}}\psi^\prime,\qquad
U\to e^{-i \gamma_5 \bbox{\tau}\cdot\bbox{ a}} \;U\; 
e^{-i \gamma_5 \bbox{\tau}\cdot \bbox{a}}.
\label{chiral}\end{eqnarray}
This invariance shows that our  scalar meson $\phi$ is not a chiral
partner of the pion.
Note the presence of the term $U(M+g_s\phi)$ 
which was incorrectly given as $MU+g_s\phi$ in Refs.~\cite{gam97a,gam97b}.

The constant $M\over f$ plays the role of the bare 
pion-nucleon coupling constant.
If $f$ is chosen to be the pion decay constant, the Goldberger-Trieman
relation says that the axial vector coupling constant
$g_A=1$. This is not really a problem because loop effects
can make up the needed 25\% effect. Corrections of that size 
are typical of order $({M\over f})^3$ effects found in the 
cloudy bag model\cite{cbm} for many observables, including $g_A$. We also note
that the  $\Delta$ is not treated as an  explicit degree of freedom
in the above Lagrangian.

The present Lagrangian
may be thought of
as a low energy effective
theory for nuclei under normal conditions. 
A more sophisticated Lagrangian is reviewed in \cite{sw97} and used in 
\cite {fu96a}; the present
one is used to show that light front techniques can be applied to
 hadronic theories relevant for nuclear physics.
This hadronic model, when evaluated in mean field approximation, 
gives\cite{bsjdw}  at least a qualitatively 
good description of many (but not all) 
nuclear properties and reactions. There are a variety of problems occurring
when higher order terms are included\cite{sw97}. The aim here is
to use a reasonably sophisticated 
Lagrangian to study the effects that one might obtain by using
a light front formulation.

Ref.~\cite{gam97b} contains the details of the quantization procedure; we 
 re-state the relevant  results here.
An essential feature is  the
quantization of spin 1/2 fermions. Although described by four-component
spinors, these fields have only two independent degrees of freedom. 
The light front formalism allows a convenient separation of
dependent and independent variables via the 
projection operators $\Lambda_\pm\equiv \gamma^0\gamma^\pm/2$\cite{des71}, with
$\psi'_\pm\equiv\Lambda_\pm \psi'_\pm$.
The independent Fermion degree of freedom
is chosen to be $\psi'_+$. The properties of the projection operators are
discussed in Appendix A.  One gets coupled equations for $\psi'_\pm$:
\begin{eqnarray}
(i\partial^--g_vV^-)\psi'_+=(\bbox{\alpha}_\perp\cdot
(\bbox{p}_\perp-g_v\bbox{V}_\perp)+\beta U (M+g_s\phi))\psi'_-\nonumber\\
(i\partial^+-g_vV^+)\psi'_-=(\bbox{\alpha}_\perp\cdot
(\bbox{p}_\perp-g_v\bbox{V}_\perp)+\beta U (M+g_s\phi))\psi'_+.
\end{eqnarray}
 The relation between $\psi'_-$ and $\psi'_+$ 
is very complicated unless one may set the plus component of the vector field
to zero\cite{lcrevs}. This is immediately obtained in QED and QCD by choosing
an appropriate gauge in which  the plus-component of the vector potential
vanishes.  Here
the non-zero mass of the vector meson prevents such a choice. 
 Instead,
one     simplifies   the equation
for $\psi'_-$ by\cite{des71,yan34}
   transforming  the Fermion field according to 
\begin{equation}
  \psi'=e^{-ig_v\Lambda(x)}\psi \label{trans}
\end{equation}
   with 
$
\partial^+ \Lambda=V^+
.$
 This 
transformation  leads to the result\cite{gam97b} 
\begin{eqnarray}
(i\partial^--g_v \bar V^-)\psi_+=(\bbox{\alpha}_\perp\cdot 
(\bbox{p}_\perp-g_v\bbox{\bar V}_\perp)+\beta U(M+g_s\phi))\psi_-\nonumber\\
i\partial^+\psi_-=(\bbox{\alpha}_\perp\cdot 
(\bbox{p}_\perp-g_v\bbox{\bar V}_\perp)+\beta U(M+g_s\phi))\psi_+\,
\label{yan}
\end{eqnarray}
where
\begin {equation}
 \partial^+\bar V^\mu=\partial^+V^\mu-\partial^\mu V^+= V^{+\mu}.
 \label{vbar}
\end{equation}
 The fields  $V^\mu$
enter the meson  
 field equations, but the fields $\bar V^\mu $ enter the fermion field
 equations. 
The eigenmode expansion  for $\bar V^\mu$ is given by:
\begin{equation}
\bar V^\mu(x)=
\int{ d^2k_\perp dk^+ \theta(k^+)\over (2\pi)^{3/2}\sqrt{2k^+}}
\sum_{\omega=1,3}\bar\epsilon^\mu(\bbox{k},\omega)\left[
a(\bbox{k},\omega)e^{-ik\cdot x}
+a^\dagger(\bbox{k},\omega)e^{ik\cdot x}\right],\label{nvfield}
\end {equation} 
where the polarization vectors $\bar\epsilon^\mu(\bbox{k},\omega)$ are
given by\cite{yan34}:
\begin{eqnarray}
\bar\epsilon^\mu(\bbox{k},\omega)=
\epsilon^\mu(\bbox{k},\omega)-{k^\mu\over k^+}\epsilon^+(\bbox{k},\omega),
\end{eqnarray}
with the properties
\begin{eqnarray}
k^\mu\bar\epsilon_\mu(\bbox{k},\omega)=-{m_v^2\over k^+}
\epsilon^+(\bbox{k},\omega),\;
\sum_{\omega=1,3}\bar\epsilon^\mu(\bbox{k},\omega)
\bar\epsilon^\nu(\bbox{k},\omega)=-(g^{\mu\nu}-g^{+\mu}{k^\nu\over k^+}
-g^{+\nu}{k^\mu\over k^+}).
\label{ebar}\end{eqnarray}

The use of the Fermion field equation allows one to obtain the 
light front Hamiltonian density
\begin{eqnarray}
T^{+-}=\bbox{\nabla}_\perp\phi\cdot\bbox{\nabla}_\perp\phi +m_\phi^2\phi^2
+{1\over 4}(V^{+-})^2+{1\over 2}V^{kl}V^{kl} +m^2_vV^kV^k\nonumber\\
+(\bbox{\nabla}_\perp\bbox{\pi})^2+ 
{\left({1\over 2}\bbox{\nabla}_\perp\pi^2\right)^2\over \pi^2}
\left(1-{f^2\over \pi^2}\mbox{sin}^2{\pi\over f}\right)+m_\pi^2f^2 
\mbox{sin}^2{\pi\over f}
\nonumber\\
+2\psi^\dagger_+\left(i {1\over 2}\stackrel{\leftrightarrow}{\partial}^-
-g_v\bar V^-\right)\psi_+.\label{tpm1}
\end{eqnarray}

The expression (\ref{tpm1}) 
is useful for situations, such as in the mean field 
approximation,  for which a simple expression
for 
$\psi_+$ is known. This is not always the case, so it is worthwhile
to use the Dirac equation to express 
$T^{+-}$ in an alternate form:
\begin{eqnarray}
T^{+-}=\bbox{\nabla}_\perp\phi\cdot\bbox{\nabla}_\perp\phi +m_\phi^2\phi^2
+{1\over 4}(V^{+-})^2+{1\over 2}V^{kl}V^{kl} +m^2_vV^kV^k\nonumber\\
+(\bbox{\nabla}_\perp\bbox{\pi})^2+ 
{\left({1\over 2}\bbox{\nabla}_\perp\pi^2\right)^2\over \pi^2}
\left(1-{f^2\over \pi^2}\mbox {sin}^2{\pi\over f}\right)+m_\pi^2f^2 
\mbox{sin}^2{\pi\over f}
\nonumber\\
+\bar\psi\left(\bbox{\gamma}_\perp
\cdot(\bbox{p}_\perp-g_v\bbox{\bar V}_\perp
)+U(M+g_s\phi)\right)\psi.\label{tpm2}
\end{eqnarray}
The relationship between $\psi$ and $\xi$ ($\xi_-$ contains no interactions)
is discussed in Ref.~\cite{gam97b}.
It is $\xi$ that is expanded in creation and
destruction operators according to
\begin{equation}
\xi(x)=\int{ d^2k_\perp dk^+ \theta(k^+)\over (2\pi)^{3/2}\sqrt{2k^+}}
\sum_{\lambda=+,-}\left[u(\bbox{k},\lambda)e^{-ik\cdot x}b(\bbox{k},\lambda)+
v(\bbox{k},\lambda)e^{+ik\cdot x}d^\dagger(\bbox{k},\lambda)\right].
\label{dq}
\end{equation}

The spinors $u(\bbox{k},\lambda)$
are the usual equal time Dirac spinors, of normalization
$\bar{u}u=2M$. It is legitimate to use these
because one is free to choose the representation of the solutions of the
Dirac equation in an infinite number of ways. In particular, 
the correct fermionic anti-commutation relation for $\xi_+$ is obtained with
these spinors \cite{yan12}.

The
Hamiltonian is a sum of a free $P^-_0(N)$ and interacting terms $P^-_I(N)$:
\begin{equation}
P_0^-(N)={1\over 2}\int d^2x_\perp dx^-
\bar\xi\left(\bbox{\gamma}_\perp\cdot\bbox{p} +M\right)\xi. \label{freef}
\end{equation}

\begin{equation}
P^-_I=v_1+v_2+v_3,  \label{defv}
\end{equation}
with 
\begin{equation}
v_1=\int d^2x_\perp dx^-\bar\xi\left(g_v\gamma\cdot\bar V+M(U-1)+g_s\phi U
\right)\xi,\label{v1}
\end{equation}
\begin{equation}
v_2=\int d^2x_\perp dx^-\bar\xi\left(-g_v\gamma\cdot\bar V
+M(U-1)+g_s\phi U\right)\;{\gamma^+\over 2p^+}\;\left(-g_v\gamma\cdot\bar V
+M(U-1)+g_s\phi U
\right)\xi, \label{v2}
\end{equation}
and 
\begin{equation}
v_3={g_v^2\over32}\int d^2x_\perp dx^-\int dy^-_1
\bar\xi(\bbox{x}_\perp,y^-_1)\gamma^+\xi(\bbox{x}_\perp,y^-_1)
\epsilon(x^--y^-_1)\int dy^-_2\epsilon(x^--y^-_2)
\bar\xi(\bbox{x}_\perp,y^-_2)\gamma^+\xi(\bbox{x}_\perp,y^-_2).\label{v3}
\end{equation}

The term $v_1$ accounts for the emission or absorption of a
 single vector or scalar meson, as well as the emission or absorption
of any number of pions through the operator $U-1$. The term $v_2$ 
includes contact terms in which there is propagation of an instantaneous 
fermion. The term $v_3$ accounts for the propagation of an instantaneous
vector meson.

The component that is related to the plus momentum is $T^{++}$.
The necessary expression is given by 
\begin{eqnarray}
T^{++}=V^{ik}V^{ik}
+m_v^2V^+ V^+
+\bar\psi\gamma^+ i\partial^+ \psi\nonumber\\
+\partial^+\phi\partial^+\phi
+\partial^+\bbox{\pi}\cdot \partial^+ \bbox{\pi}
+\bbox{\pi}\cdot\partial^+\bbox{\pi}
{\bbox{\pi}\cdot\partial^+\bbox{\pi}\over\pi^2}(1-{f^2\over \pi^2}
\mbox{sin}^2{\pi\over f}). \label{tpp}
\end{eqnarray}

\section{Nucleon-Nucleon Scattering via One-Boson Exchange Potentials}

The correlations between nucleons are caused by the nucleon-nucleon
interaction. Thus a necessary first step towards a light-front
theory of nuclear
correlations is the derivation of a light-front theory of the nucleon-nucleon
interaction. Previous work \cite{gam97a,gam97b} showed that the light-front
version of the Lippmann-Schwinger equation, the Weinberg equation, can be
transformed (with one difference remaining\cite{difference})
into the Blankenbecler-Sugar equation.
Kinematic  invariance 
under boosts in the three-direction is maintained, and
 we shall
obtain a one-boson exchange potential which is in reasonably good
agreement with the NN phase shifts.

\subsection{General Formalism}
It is worthwhile to begin by reviewing how using  the light-front Hamiltonian
of Eqs.~(\ref{freef}-\ref{v3}) leads to the one-boson exchange potential.
This derivation is  useful in understanding how the full nuclear wave
function discussed in Sect.~VI is related to the nucleonic truncation of
Sect.~IV. Consider the scattering process $1+2\to 3+4$.
The use of second-order perturbation theory shows that the
lowest-order contribution to the nucleon-nucleon scattering
amplitude is given by
\begin{equation}\langle 3,4|
 K|1,2\rangle=\langle 3,4| v_1g(P_{ij}^-)v_1+v_3|1,2\rangle,
 \label{kdef0}
 \end{equation}
 with
\begin{equation}
  g_0(P_{ij}^-)\equiv {1\over P_{ij}^--P_0^-} \; , \label{gij}
  \end{equation}
  where $P_{ij}^-$ is the negative component of the total initial or final
  momentum which are the same.
  In 
  constructing the NN potential, one uses conservation of four-momentum between
  the initial and final NN states.
  The  expression (\ref{kdef0}) yields a 
 one-boson exchange approximation to the nucleon-nucleon potential.

 It is worthwhile to discuss  the energy denominator $ P_{ij}^--P_0^-$
 in more detail. To be specific, suppose that $k_1^+>k_3^+$. Then
 the emitted meson of mass $\mu$
 has momentum $k$ with  $k^+=k_1^+-k_3^+, \bbox{k}_\perp=
 {\bbox{k}_1}_\perp-{\bbox{k}_3}_\perp$ and $k^-={k_\perp^2+\mu^2\over k^+}$.
 Then
 \begin{equation}
    P_{ij}^--P_0^- =  P_{12}^--P_0^- = P_{34}^--P_0^- =(k^-_1-k_3^-)
    -{{k_\perp}^2+\mu^2\over k_1^+-k^+_3}.
 \label{pdiff}   \end{equation}
    The interaction $K$ also contains a factor of $k^+$ in the denominator, so
    that the relevant denominator is $D\equiv k^+ ( P_{ij}^--P_0^- )=
    (k_1^+-k_3^+) (k_1^--k_3^-)-k_\perp^2-\mu^2=q^2-\mu^2$. This last familiar
    form involves the four-momentum transfer between nucleons 1 and 3
    ($k_1^+>k_3^+) $ and leads to the usual Yukawa-type potentials. It is
    also useful to explore the form of the energy denominator using light-front
    variables by first defining the plus-component,$P^+$, of
    the initial and final total momentum. We
    may also
    define $k_1^+=xP^+$ and $k_3^+=x'P^+$ in which $x,x'$ are invariant under
    Lorentz transformations  in the three-direction. Then using 
    (\ref{pdiff}), we find
    \begin{equation}
      D={{{k_1}_\perp}^2 +M^2\over x}-{{{k_3}_\perp}^2 +M^2\over x'} -
      k_\perp^2-\mu^2. \label{deq}
      \end{equation}
      This quantity is also    invariant under
    Lorentz transformations  in the three-direction.

 A straightforward evaluation of Eq.~(\ref{kdef0})
 using Eqs.~(\ref{v1}-\ref{v3}) leads to the result 
\begin{equation}\langle 3,4|
 K|1,2\rangle=2\;\langle 3,4|V|1,2\rangle\quad
 {M^2\delta^{(2,+)}(P_i-P_f)\over \sqrt{k_1^+k_2^+k_3^+k_4^+}},\label{kdef}
\end{equation} 
where $\delta^{(2,+)}(P_i-P_f)\equiv
\delta^{(2)}(\bbox{P}_{i\perp}-\bbox{P}_{f\perp}) \delta(P_i^+-P_f^+)$ and $V$
is the standard expression for the sum of the $\pi,\phi $ and vector meson
exchange potentials:
\begin{equation}\langle 3,4|V|1,2\rangle=\langle 3,4|V(\phi) +V(\bbox{\pi})+
  V(V)|1,2\rangle. 
\end{equation}
The operator $K$ is twice the usual two-nucleon potential times a factor which
includes  the light front phase space factor and a momentum-conserving delta
function. Note that $dk^+/k^+=dk^3/E(k)$ for free nucleons where
$k^+=E(k) +k^3$.

For the exchange of
scalar and pseudoscalar mesons, only the term $v_1g_0(P_i^-)v_1$
enters, and one finds
\begin{equation}
\langle 3,4|V(\phi,\pi)|1,2\rangle
 = {\bar u(4)\Gamma u(2)\; \bar u(3)\Gamma u(1) \over
4M^2 (2\pi)^3 \left(q^2-\mu^2\right)} \; , \label{sex}
\end{equation}
in which the momentum transfer $q$ is given by
\begin{equation}
  q\equiv k_3-k_1.
  \label{q}
  \end{equation}
The notation is that $u(i)$ is the Dirac-spinor 
for a free nucleon of quantum numbers $i$, and $\Gamma$ is either of the
form $g_s$  or $i\;g_\pi\gamma_5{\bbox{\tau}}$. 
The derivation of the contribution of vector meson exchange proceeds by 
including the meson exchange $v_1g_0(P_i^-)v_1$
plus the meson instantaneous term $v_3$, and
the result 
takes the familiar form:
  \begin{equation}
\langle 3,4|
V(V)|1,2\rangle = -g_v^2
{\bar u(4)\gamma_\mu u(2) \bar u(3)\gamma^\mu u(1) \over
4M^2 (2\pi)^3 \left(q^2-m_v^2\right)}.\label{vex}
\end{equation}
The expressions  (\ref{sex}) and (\ref{vex}) represent
the usual \cite{rm,Mac89,Mac93,rm1,geb} expressions for  the chosen
one-boson exchange 
potentials, if no form 
factor effects are included.
The sum
of the amplitudes arising from each of the 
individual one-boson exchange terms
gives  the invariant amplitude to second order in each of the
coupling constants. 
The factors $1\over 4M^2$ in Eqs.~(\ref{sex})   and (\ref{vex})
can be thought of as renormalizing the spinors so that $\bar{u}u=1$, and
the factors $\sqrt{M\over k^+}$ of Eq.~(\ref{kdef}) serve 
to further change the normalization to $u^\dagger u=1$

These amplitudes are strong, so computing the nucleon-nucleon scattering 
amplitude and phase shifts requires including higher order terms.
One may include  a sum which gives unitarity by including 
all iterations of the scattering operator $K$ 
through intermediate two-nucleon states:
\begin{equation} 
\langle3,4|{\cal M}|1,2\rangle
=\langle 3,4|
K|1,2\rangle+
\sum_{\lambda_5,\lambda_6}\int  \langle 3,4|
K|5,6\rangle 
{2M^2\over p_5^+p_6^+}
{d^2p_{5\perp}dp^+_5\over P_i^--(p_5^-+p_6^-)+i\epsilon}
\langle5,6|{\cal M}|1,2\rangle. \label{weinberg}
\end{equation}
The factor $ {M^2\delta^{(2,+)}(P_i-P_f)\over \sqrt{k_1^+k_2^+k_3^+k_4^+}}$
appears in both  of the terms on the right-hand-side
of Eq.~(\ref{weinberg}), so it is worthwhile
to define a $T$-Matrix $T$ using
\begin{equation}
  {\cal M}\equiv 2
  T {M^2\delta^{(2,+)}(P_i-P_f)\over \sqrt{k_1^+k_2^+k_3^+k_4^+}}.
\end{equation}
One realizes that Eq.~(\ref{weinberg})
is of the form of the Weinberg equation\cite{We66} (see Ref.~\cite{fs1})
by expressing the plus-momentum   variable in terms of a light-front 
momentum  fraction
$\alpha$ such that
\begin{equation}
p_5^+=\alpha P_i^+,\label{alphadef}
\end{equation}
and using the relative and total momentum variables:
\begin{eqnarray}
\bbox{p}_\perp\equiv (1-\alpha)\bbox{p_5}_\perp-\alpha
\bbox{p_6}_\perp \; , \nonumber\\
\bbox{P_i}_\perp=\bbox{p_5}_\perp+\bbox{p_6}_\perp \; . \label{alpha2}
\end{eqnarray}
Then,
\begin{equation} 
\langle3,4|
T|1,2\rangle
=\langle 3,4|
V|1,2\rangle+
\int\sum_{\lambda_5,\lambda_6} \langle 3,4|
V|5,6\rangle 
{2M^2\over \alpha(1-\alpha)}
{d^2p_\perp d\alpha\over P_i^2-{p_{\perp}^2+M^2\over\alpha(1-\alpha)}
+i\epsilon}
\langle5,6|T
|1,2\rangle, \label{419}
\end{equation}
where $P_i^2$ is the square of the total initial  four-momentum,
otherwise known as the invariant energy $s$ and 
${p_{\perp}^2+M^2\over\alpha(1-\alpha)}$
 is the corresponding quantity for the intermediate state.
 Because the kernal $V$ 
is itself  invariant under Lorentz transformations in the  three-direction
and the integral involves $p_\perp $ and $\alpha$ 
the procedure of solving this equation gives  $T$ with the same invariance. 

Equation~(\ref{419}) can
in turn be re-expressing as the Blankenbecler-Sugar (BbS) equation
\cite{BbS} 
by  using the variable transformation\cite{Te 76}:
\begin{equation}
\alpha={E(p)+p^3\over 2E(p)}, \label{alpha}
\end{equation}
with $E(p)\equiv\sqrt{\bbox{p}\cdot\bbox{p}+M^2}$.
The result is:
\begin{equation}
\langle3,4|T
|1,2\rangle
=\langle3,4|V
|1,2\rangle+\int
\sum_{\lambda_5,\lambda_6} \langle 3,4|V
|5,6\rangle
{M^2\over E(p)}
{d^3p \over p_i^2-p^2
+i\epsilon}
\langle5,6|T
|1,2\rangle, \label{bsbs}
\end{equation}
which is the desired equation. Rotational invariance is manifestly obeyed. 
The three-dimensional propagator is exactly that of the BbS equation. 
There is,
one difference between Eq.~(\ref{bsbs}) and the standard BbS equation.
 Our one-boson exchange potentials depend on the 
square of the four momentum $q^2$ transferred when a meson is absorbed or
emitted by a nucleon. Thus the energy difference between the initial and final
on-shell nucleons is included and $q^0\ne 0$. This non-zero value is a
consequence of the invariance of $D$ of Eq.~(\ref{deq}) under Lorentz
transformations in the three-direction.  
 The usual derivation of the BbS equation from
the Bethe-Salpeter equation specifies that  $q^0=0$ is used 
in the meson propagator. Including $q^0 \ne 0$ instead of $q^0=0$ increases the
range of the potential relative to the usual treatment,
and its consequences are explored below.
One can convert Eq.~(\ref{bsbs}) into
the Lippman-Schwinger equation of non-relativistic scattering theory
by removing the factor $M/E(p)$ with a simple transformation\cite{pl}.

\subsection{Generation of a Realistic One-Boson Exchange Potential}
The present results are that one can use the light front technique to
derive nucleon-nucleon potentials in the one-boson exchange (OBE)
approximation and use these in an appropriate wave equation.
 Our purpose here is to show  that the present procedure 
yields potentials essentially identical to the Bonn OBEP potentials
\cite{Mac89,Mac93}  and
these potentials 
lead to a good description of the NN data.

The Bonn one-boson exchange  potentials employ six different mesons, namely,
$\pi,\eta, \omega,\rho, \sigma$ and the (isovector scalar)
$\delta/a_0$ meson.
The present formalism can account   for the
$\pi,\eta, \omega$ and $ \sigma$ in an approximately chiral invariant manner.
We wish to add in couplings $\bar{\psi}\bbox{\tau}\cdot\bbox{\delta}\psi$
  and $\bar{\psi}\bbox{\tau}\cdot\bbox{\rho}^\mu\gamma_\mu\psi$
 in a chiral invariant manner. Simply adding such terms to the Lagrangian
 of Eq.~(\ref{lag}) would lead to a violation of the approximate symmetry of
 Eq.~(\ref{chiral}). 
 However, one can redefine the operator $U$ so that
 the symmetry remains. We replace the operator $\bar{\psi}'U\psi'$ in the
 Lagrangian (\ref{lag}) by $\bar{\psi}'\tilde {U}\psi'$ :
 \begin{equation}
   \tilde{U}\equiv
   e^{{i\over 2 f_\rho}\bbox{\tau}\cdot\bbox{\rho}^\mu\gamma_\mu}
     e^{{i\over 2 f_\delta}\bbox{\tau}\cdot\bbox{\delta}}U
        e^{{i\over 2 f_\delta}\bbox{\tau}\cdot\bbox{\delta}}
          e^{{i\over 2 f_\rho}\bbox{\tau}\cdot\bbox{\rho}^\mu\gamma_\mu}.
            \end{equation}
Then the new  Lagrangian  is invariant under the transformation
\begin{eqnarray}
\psi^\prime\to e^{i \gamma_5 \bbox{\tau}\cdot\bbox{a}}\psi^\prime,\qquad
\tilde{U}\to e^{-i \gamma_5 \bbox{\tau}\cdot\bbox{ a}} \;U\; 
e^{-i \gamma_5 \bbox{\tau}\cdot \bbox{a}}.
\label{nchiral}\end{eqnarray} In the present application 
we expand the exponential to first order in the meson fields.

The final term we need to include is the 
 tensor 
$\sigma_{\mu\nu}q^\nu$ part of the $\rho$-nucleon interaction. 
The presence of such a tensor interaction makes it difficult (or
impossible) to write the equation for $\psi_-$ as $\psi_-=1/p^+\cdots
\psi_+.$ This is a possible problem
because the standard value of the ratio of
the tensor to vector $\rho$-nucleon coupling $f_\rho/g_\rho$ is 6.1,
based upon Ref.~\cite{hp}.  Reproducing the
observed values of $\varepsilon_1$ and P-wave wave phase shifts
requires a large value 
$f_\rho/g_\rho$; see Ref.~\cite{brm}.  However our Lagrangian
compensates for its lack of a $\rho$-N interaction with tensor
coupling by generating such a term via vertex correction diagrams
(which are the origin of the anomalous magnetic moment of the electron
in QED). Such diagrams might  not generate the
phenomenologically required values of the coupling constants, but all
that is needed here is that terms of the correct form be
produced. This is because the standard procedure is to choose the
values of the coupling constants so as to yield a good description of
the NN scattering data.  
Thus we simply add in the necessary tensor terms.   

This brings us to the treatment of divergent terms in our procedure.
The definition of any effective Lagrangian
requires the specification of such a procedure.  For the present, 
it is sufficient to say that we introduce form factors, $F_\alpha(q^2)$
which reduce the strength of the $\alpha$ meson-nucleon coupling for
large values of $-q^2$. This is also the procedure of 
Refs.~\cite{Mac89,Mac93,rm1}. In principle, calculating the
higher order terms using the correct Lagrangian can lead to consistent
calculations of these form factors.   We use a more
phenomenological approach here.

The net result is that the one-boson exchange treatment of the
nucleon-nucleon potential and the T-matrix resulting from its use in
the BbS equation is essentially the same as the one-boson exchange
procedure of Refs.~\cite{rm,Mac89,Mac93,rm1}. The only difference is
the keeping
of the retardation effects---the square of the four-vector momentum transfer
enters in our potentials.

\subsection{Specific  One-Boson-Exchange Amplitudes}
The above formalism yields
a one-boson-exchange potential (OBEP)
which is  a sum of
one-particle-exchange amplitudes of certain bosons
with given mass and coupling. Our explicit expressions are presented here.
As noted above, we use the six non-strange bosons with masses
below 1 GeV/c$^2$. Thus,
\begin{equation}
V_{OBEP}=\sum_{\alpha=\pi,\eta,\rho,\omega,a_0,\sigma}
V^{OBE}_{\alpha}
\label{VOBEP}
\end{equation}
with $\pi$ and $\eta$ pseudoscalar (ps),
$\sigma$ and $a_0/\delta$ scalar (s), and
$\rho$ and $\omega$ vector (v) particles.

The  
OBE amplitudes (which are the contributions to $V$ of our formalism)
in the two-nucleon center-of-mass (c.m.) frame are given by :
\begin{eqnarray}
\lefteqn{\hspace{-1cm}\langle {\bbox k'} \lambda_{3}\lambda_{4}|V^{OBE}_{ps}|
{\bbox k}\lambda_{1}\lambda_{2}\rangle} \nonumber \\
 & = &
-\frac{{g^{2}_{ps}}}{(2\pi)^3 4M^2}
\bar{u}({\bbox k'},\lambda_{3})  \gamma^{5} u({\bbox k},\lambda_{1})
\bar{u}({\bbox -k'},\lambda_{4})  \gamma^{5} u({\bbox -k},\lambda_{2})
 [{q}^{2}-m_{ps}^{2}]^{-1} \; ;
\label{VOBEpv}\\
\nonumber \\
\lefteqn{\hspace{-1cm}\langle {\bbox k'} \lambda_{3}\lambda_{4}|V^{OBE}_{s}|
{\bbox k}\lambda_{1}\lambda_{2}\rangle}\nonumber \\
 & = & \frac{g^{2}_{s}}{(2\pi)^3 4M^2}
\bar{u}({\bbox k'},\lambda_{3})    u({\bbox k},\lambda_{1})
\bar{u}({\bbox -k'},\lambda_{4})   u({\bbox -k},\lambda_{2})
 [{q}^{2}-m_{s}^{2}]^{-1} \; ;
\label{VOBEs}\\
\nonumber \\
\lefteqn{\hspace{-1cm}\langle {\bbox k'} \lambda_{3}\lambda_{4}|V^{OBE}_{v}|
{\bbox k}\lambda_{1}\lambda_{2}\rangle}\nonumber \\
 &  = &
-\frac{1}{(2\pi)^3 4M^2} \{g_{v}
\bar{u}({\bbox k'},\lambda_{3})  \gamma_{\mu} u({\bbox k},\lambda_{1})
+\frac{f_{v}}{2M}
\bar{u}({\bbox k'},\lambda_{3})
\sigma_{\mu\nu}i(k_3-k_1)^{\nu}
 u({\bbox k},\lambda_{1})\}
\nonumber \\ & & \times
\{g_{v}\bar{u}({\bbox -k'},\lambda_{4})
  \gamma^{\mu} u({\bbox -k},\lambda_{2})
+\frac{f_{v}}{2M}\bar{u}({\bbox -k'},\lambda_{4})
\sigma^{\mu\nu}i(k_4-k_2)_{\nu}
   u({\bbox -k},\lambda_{2})\}
\nonumber \\ 
& & \times [{q}^{2}-m_{v}^{2}]^{-1}
\nonumber \\ & = &
-\frac{1}{(2\pi)^3 4M^2} \{(g_{v}+f_{v})
\bar{u}({\bbox k'},\lambda_{3})  \gamma_{\mu} u({\bbox k},\lambda_{1})
-\frac{f_{v}}{2M}
\bar{u}({\bbox k'},\lambda_{3})
(k_3+k_1)_{\mu}
 u({\bbox k},\lambda_{1})\}
\nonumber \\ & & \times
\{(g_{v}+f_{v})\bar{u}({\bbox -k'},\lambda_{4})
  \gamma^{\mu} u({\bbox -k},\lambda_{2})
-\frac{f_{v}}{2M}\bar{u}({\bbox -k'},\lambda_{4})
(k_4+k_2)^{\mu}
   u({\bbox -k},\lambda_{2})\}
\nonumber \\  &  & \times [{q}^{2}-m_{v}^{2}]^{-1} \; .
\label{VOBEv}
\end{eqnarray}
Our notation in the c.m.\ frame is such that 
in-coming nucleon 1 carries helicity $\lambda_1$ 
and four-momentum $k_1=k=(E,{\bbox k})$ 
with $E \equiv \sqrt{M^2+{\bbox k}^2}$, and
in-coming nucleon 2 carries helicity $\lambda_2$ 
and four-momentum $k_2=(E,-{\bbox k})$;
the out-going nucleons have
$\lambda_3$, 
$k_3=k'=(E',{\bbox k'})$ 
with $E' \equiv \sqrt{M^2+{\bbox k'}^2}$, and
$\lambda_4$, 
$k_4=(E',-{\bbox k'})$ .
The square of the four-momentum transfer between
the two nucleons is 
$q^2=(k_3-k_1)^2=(k'-k)^2=(E'-E)^2-({\bbox k'}-{\bbox k})^2$.
 The Gordon identity~\cite{Bj} is used in the 
evaluation of the tensor coupling~\cite{foot1}.
For the isospin-vector bosons $\pi$, $a_0$ and $\rho$,
the above amplitudes must be multiplied by
{\boldmath $\tau_{1} \cdot \tau_{2}$}.

With an eye on the nuclear matter calculations to be conducted
later in this paper, we note that in the factor $f_v/2M$ of the
tensor coupling the nucleon
mass $M$ is used as a scaling mass to make the coupling constant
$f_v$ dimensionless. This scaling mass could be anything.
Therefore, this $M$ is not to be replaced by $M^*$ in the
nuclear medium.

In this subsection, we use
Dirac spinors (in helicity representation) given by
\begin{eqnarray}
u({\bbox k},\lambda_1)&=&\sqrt{{E+M}}
\left( \begin{array}{c}
       1\\
       \frac{2\lambda_1 |{\bbox k}|}{E+M}
       \end{array} \right)
|\lambda_1\rangle
\label{FreeSpinor} \\
u(-{\bbox k},\lambda_2)&=&\sqrt{E+M}
\left( \begin{array}{c}
       1\\
       \frac{2\lambda_2 |{\bbox k}|}{E+M}
       \end{array} \right)
|\lambda_2\rangle
\end{eqnarray}
with
\begin{equation}
|\lambda_1\rangle= \chi_{\lambda_1}\; , \;\;\;\;
|\lambda_2\rangle= \chi_{-\lambda_2}\; ,
\end{equation}
where $\chi$ denotes the conventional Pauli spinor.
The normalization is
$\bar{u}({\bbox k},\lambda) u({\bbox k},\lambda)=2M$.

At each meson-nucleon vertex, a form factor is applied which has the 
analytic form
\begin{equation}
{ F}_\alpha(q^2)=\left( \frac{\Lambda^2_\alpha-m^2_\alpha}
{\Lambda^2_\alpha-q^2}\right)^{n_\alpha},
\label{Cutoff1}
\end{equation}
with $m_\alpha$ the mass of the meson involved and
$\Lambda_\alpha$ the so-called cutoff mass; 
$n_\alpha=1$ for pseudoscalar and scalar mesons and
$n_\alpha=2$ for vector mesons.
Thus, the OBE amplitudes Eqs.~(\ref{VOBEpv})-(\ref{VOBEv}) are all multiplied
by ${ F}_\alpha^2$.

\subsection{Two-Nucleon Scattering}
In the two-nucleon c.m.\ frame, the scattering amplitude ${ T}$ 
is the solution of the integral equation
\begin{equation}
{ T}({\bbox k'},{\bbox k})=V({\bbox k'},{\bbox k})+
\int d^{3}p
V({\bbox k'},{\bbox p})
\frac{M^{2}}{E_{\bbox p}}
\frac{1}
{{\bbox k}^{2}-{\bbox p}^{2}+i\epsilon}
{ T}({\bbox p},{\bbox k})
\label{TEQ}
\end{equation}
where ${\bbox k}$, ${\bbox p}$ and ${\bbox k'}$ are
the initial, intermediate and final relative momenta, respectively,
of the two interacting nucleons; and
$E_{\bbox p}\equiv \sqrt{M^2+{\bbox p}^2}$.
This is equation ~(\ref{bsbs}) with the spin indices suppressed for the
purpose of simplicity.
The corresponding equation for the $K$-matrix (which we denote by $R$)
is
\begin{equation}
R({\bbox k'},{\bbox k})=V({\bbox k'},{\bbox k})+
{\cal P} \int d^{3}p
V({\bbox k'},{\bbox p})
\frac{M^{2}}{E_{\bbox p}}
\frac{1}
{{\bbox k}^{2}-{\bbox p}^{2}}
R({\bbox p},{\bbox k})
\label{REQ}
\end{equation}
where ${\cal P}$ denotes the principal value integral.

Using standard techniques~\cite{Mac93,foot1}, 
the potential and the scattering equation
are decomposed into partial waves. Numerical solutions
are obtained by the matrix inversion method~\cite{Mac93,HT70}.
For an uncoupled partial wave,
phase shifts are then derived from the on-shell $K$-matrix by
\begin{equation}
\tan\: \delta^J(T_{lab}) = -\frac{\pi}{2} 
|{\bbox k}|
\frac{M^2}{E}
\: R^J
(|{\bbox k}|,
|{\bbox k}|)
\label{PhaseRel}
\end{equation}
with $T_{lab}=2{\bbox k}^2/M$
and $J$ the total angular momentum of the partial-wave state.
For coupled partial waves and other technical details, see
Ref.~\cite{Mac93}.

\subsection{Results for the two-nucleon system}
Following established procedures~\cite{Mac89,Mac93}, the coupling constants
and cutoff masses of the six OBE amplitudes are varied within 
reasonable limits such as to reproduce the two-nucleon bound
state (deuteron) and the two-nucleon scattering data below the
inelastic threshold (about 300 MeV laboratory kinetic energy).

In Table~I, we show the meson parameters for our newly
constructed Light-Front (LF) OBEP together with the predictions
for the deuteron as well as low-energy neutron-proton ($np$)
scattering.
For comparison, we also give the parameters from an OBEP that
was previously constructed and applied in the Dirac-Brueckner
approach to nuclear matter~\cite{Mac89,rm1}. The latter
uses the Thompson formalism~\cite{Tho70} which is very similar
to the BbS formalism---the propagator in Eq.~(\ref{TEQ}) 
contains an extra factor of $M/E_p$.
Note that the Thompson OBEP uses $n_\alpha = 1$
also for vector meson form factors, which explains the differences
in the vector meson cutoff masses between the two OBEP.

Phase shifts for $np$ scattering are shown in Fig.~1 for
all partial waves with $J\leq 2$.
Over-all, the reproduction of the $NN$ data by our LF OBEP is quite
satisfactory and certainly as good as by
OBEP constructed within alternative relativistic
frameworks.
Based upon these results, we feel confident in applying
this OBEP to the relativistic nuclear many-body problem.

\section{Nucleonic Truncation For The Many-Body Problem}

Now that the light front treatment of nucleon-nucleon scattering is in
hand, we may proceed to the problem of computing the properties of infinite
nuclear matter.
We  derive a light front
Brueckner
theory from first principles  starting with the field-theoretic
light-front Hamiltonian. 

The nuclear wave function for the ground state of infinite nuclear matter 
at rest is
 defined as $|\Psi\rangle$, and we wish to solve the equation
\begin {equation}
P^-|\Psi\rangle=M_A|\Psi\rangle, \label{sc}
\end{equation} 
in which 
$P^-$ is the light-front Hamiltonian
of 
Eqs.~(\ref{tpm1})-(\ref{v3}). For a nuclear system at rest we must have also
the result that 
\begin {equation}
P^+|\Psi\rangle=M_A|\Psi\rangle. \label{scp}
\end{equation} 
It is necessary to discuss  the light front Hamiltonian, and to  find good
 approximate solutions of 
 the above equations.

We recall that 
\begin{equation}
 P^- = P^-_0(N) + J \label{op}
\end{equation}
in which $P^-_0(N)$ is the kinetic contribution to the $P^-$ operator,
 giving ${p_\perp^2 + m^2}\over{p^+}$ for the minus-momentum of free
 fermions.
The operator $J$ is the sum of three terms of Eqs.~(\ref{v1})-(\ref{v3}):
\begin{eqnarray}
J \equiv v_1 + v_2 + v_3 
\end{eqnarray}
The operator 
$v_1 $ gives all of the single meson-nucleon vertex functions. The operator
$v_2 $ accounts for instantaneous fermion exchanges: meson emission followed
by instantaneous fermion propagation (propagator is ${\gamma^+\over 2 p^+}$)
followed by another meson emission. The operator $v_3 $ accounts for the 
instantaneous propagation of vector mesons.

We shall proceed towards an approximate solution of Eq.~(\ref{sc}), in two
stages. We shall first
consider the nucleons only part of the Hilbert space. This
involves
the assumption that using  a nucleon-nucleon
interaction  $K$ accounts for the meson-nucleon dynamics.
This assumption is relaxed in Section VI, which displays the formalism
necessary  to
construct the best possible potential and how to include meson degrees
of freedom  in the wave function.
Our  Hamiltonian ($P^-$) contains no terms in which the
vacuum can spontaneously
emit particles. This simplifying feature causes the
 derivations  to look
very similar to those of non-relativistic theory,
even though the treatment is
relativistic.

\subsection{Introducing the two-nucleon force}
At present all of the interactions are expressed in terms of the
meson-nucleon vertex functions and contact terms represented by the
operator $J$. We shall follow the
traditional path of using a two-nucleon potential and temporarily
eliminate the meson degrees of freedom.  One way  to accomplish this 
is to 
add  and subtract  the two-nucleon potential
to the Hamiltonian and treat terms involving the difference between $J$ and
the two-nucleon potential as a perturbation.
The use of light front dynamics mandates that we
perform this operation on the Lagrangian  because the construction of
the Hamiltonian  uses the  field equations to identify the
dynamical degrees of freedom, such as $\psi_+$.
Therefore  we need to study the effective Lagrangian:
\begin{equation}
 {\cal L_V} \equiv\bar{\psi}(i\gamma\cdot\partial -M)\psi -{{{\cal K}\over 2}}.
\label{lagv}\end {equation} 
which removes the meson-nucleon interaction term $J$ from the Lagrangian and
replaces it with the density ${{\cal K}}$
for the  two-nucleon interaction $K$ of the previous section.
This means that
\begin{equation}
  K={1\over 2}\int d^2x_\perp dx^-{\cal K},
  \end{equation}
  where $K$ is given in Eq.~(\ref{kdef}) and is twice the nucleon-nucleon
  potential times a kinematic factor.
We recall that
 one must eliminate, using  Eq.~(\ref{trans}),
the components of interaction ${{\cal K}}$
that connect $\psi_-$ to $\psi_-$.

Given the new Lagrangian ${\cal L_V}$ we may construct
the corresponding $P^\pm$ operators using the canonical definition
\begin{equation}
T^{\mu\nu}_{\cal V}=-g^{\mu\nu}{\cal L}_{\cal V}
+\sum_r{\partial{\cal L}_{\cal V}\over\partial
  (\partial_\mu\phi_r)}\partial^\nu\phi_r,\label{tvmunu}
\end {equation}
in which the degrees of freedom $\bar {\psi} $ and $\psi$
are labelled by $\phi_r$.
The term ${{\cal V}}$ involves only nucleon fields,
not their derivatives, so the second term of the energy-momentum tensor
do not enter  in
computing $T^{\mu\nu}_{\cal V}$.
The element  $\mu\nu=+-  (g^{+-}=2)$ is needed 
to construct the relevant  P-minus operator and we find
\begin{eqnarray}  
T^{+-}_{\cal V}=-2\bar{\psi}\left(i\gamma\cdot\partial -M\right)\psi
+{{\cal K}}+2\psi^\dagger_+ i\partial^-\psi_+ . \label{tv}
\end{eqnarray}
The origin
of the factor $1\over 2$ that multiplies ${\cal K}$ in the Lagrangian
(\ref{lagv}) is that  ${\cal K}$ enters here in $T^{+-}_{\cal V}$. 
It is worthwhile to  define the 
P-minus operator obtained by using ${\cal L_V} $
as $P^-_0$ with : 
\begin{eqnarray}
P^-_0 &=&{1\over 2}\int dx^-d^2x_\perp T^{+-}_{\cal V}\equiv  P^-_0(N) +K.
\label{p0m}\end{eqnarray}                                
The complete P-minus operator is given by 
\begin{eqnarray}P^- &=& P^-_0 + H_1,
 \end{eqnarray} with
 \begin{eqnarray}
 H_1 &\equiv& J-K +P^-_0(m)
\end{eqnarray}
where $P^-_0(m)$ accounts for the non-interacting mesonic contribution
to $P^{-}$. 
The formal problem of choosing  the best $ {K}$
by minimizing the effects of
$H_1=J-K$ is discussed in the Section VI. We shall assume here
that the present OBEP is a reasonably satisfactory version of the best
interaction, and we shall ignore the influence of the term $H_1$ in
calculations of the energy.

The purely nucleonic   part of the full wave function is defined as 
$\mid \Phi  \rangle$, and is
the solution of the light front Schroedinger equation
\begin{equation} P^-_0 \mid \Phi \rangle
  =\left( P^-_0(N) +K\right)\mid \Phi \rangle=
M_0 \mid \Phi \rangle  .\label{big}
\end{equation}
The eigenvalue problem stated above is considerably simpler than the initial
one, but does contain the full complications of the
nuclear many-body problem. We shall next discuss the light front Hartree-Fock
and
Brueckner Hartree-Fock \cite{bhf} approximations.

\subsection{Light Front Hartree Fock Approximation}
The equation (\ref{big}) represents a difficult many-body problem. But the
similarity between the light-front and equal-time results
obtained for the nucleon-nucleon potential indicate
that the same  physical concepts are relevant, independent of the dynamical
scheme. 
Therefore  we use 
 a scheme analogous
to that of traditional Brueckner theory.  The first step is to
introduce 
 a mean-field ($MF$) potential
 \begin{eqnarray}
 \widehat{U}&=&{1\over2}\int d^2x_\perp dx^-\bar{\psi}(x)
 \left(U_S(x)+\gamma\cdot \bar{U}_V(x)\right)\psi (x)\nonumber\\
&=&{1\over2}\int d^2x_\perp dx^- \bar{\psi}(x)\left(U_S(x)+{\gamma^+\over
    2}U_V^-(x)\right)\psi(x),\label{mf} 
\end{eqnarray}
a single-nucleon operator, with the second equation true for infinite nuclear
matter in which the only non-vanishing component of $U_V^\mu$ is
$U_V^0=U_V^-=U_V^+=\bar{U}_V^-$.
All quantities in the integral of Eq.~(\ref{mf}) are
evaluated at the same  value of $x^+$, chosen to be 0.
The operator $\widehat{U}$ is 
to be determined ultimately by the light-front G-matrix defined
below. The idea is that $\widehat{U}$ can be chosen so as to provide
a good representation of the effects of the two-nucleon interaction
$ K$. The mean field  Lagrangian 
 ${\cal L}_{MF}$ is defined by removing the effects of $ K$ and replacing
 these by the effects of $\widehat U$. Therefore we may specify 
\begin{eqnarray}
{\cal L}_{MF}\equiv\bar{\psi}\left(i\gamma_\mu
\partial^\mu-M\right)\psi -\bar{\psi}(x)\left(U_S(x)+{\gamma^+\over
    2}U_V^-(x)\right)\psi(x). \label{lmf}
\end{eqnarray}  
This Lagrangian leads  to the nucleon field equation:
\begin{eqnarray}
(i\partial^- - {U}_V^-)\,\,\, \psi_+
= (\bbox{\alpha}_\perp \cdot\bbox{p}_\perp + \beta 
(M + U_S)) \psi_- \label{lfmf}\nonumber\\
i\partial^+ \psi_-
= (\bbox{\alpha}_\perp \cdot \bbox{p}_\perp
+ \beta (M + U_S)) \psi_+ ,\label{mffe}
\end{eqnarray}
in which we have made the mean-field version of the transformation
(\ref{trans})
  with
  \begin{equation}
    \partial^+g_v\Lambda_{MF}=U_V^+(x).\label{transmf}
  \end{equation}  

  The light-front  Hamiltonian density $T_{\cal V}^{+-}$ can now be obtained
from Eq.~(\ref{tv})
  using the field equation (\ref{mffe}) as
\begin{eqnarray}  
T^{+-}_{\cal V}=2\bar{\psi}_+ i\partial^-\psi_+
-2\bar{\psi}\left({\gamma^+\over 2}{U}_V^-+U_S\right)\psi
+{{\cal K}}. \label{tvm}
\end{eqnarray}
It is also worthwhile to obtain the plus-momentum density  $T_{\cal V}^{++}$ 
which is
\begin{eqnarray}  
T^{++}_{\cal V}=2\bar{\psi}_+i\partial^+\psi_+. \label{tvp}
\end{eqnarray}

The purpose of
introducing the mean field approximation is that the eigenvalues
and eigenvectors of light-front mean field Hamiltonian
$P^-_{MF}$ are easy to obtain  and can be chosen so as best approximate the
effects of the two-nucleon interaction. The mean field
light front Hamiltonian density $T_{MF}^{+-}$ is obtained from
Eq.~(\ref{tvmunu}) using Eq.~(\ref{lmf}) (or by setting $\cal K$ to zero in
Eq.(\ref{tvm}))  as
\begin{equation}
T^{+-}_{MF}=2\bar{\psi}_+ i\partial^-\psi_+
-2\bar{\psi}\left({\gamma^+\over 2}{U}_V^-+U_S\right)\psi
, \label{tm1}
\end{equation}
and its volume integral 
is $P^-_{MF}$:
\begin{equation}
P^-_{MF}= {1\over 2}\int d^2x_\perp dx^-  T^{+-}_{MF},\label{tm2}
\end{equation}
a single-nucleon operator. Setting  $\cal K$ to zero in  Eq.~(\ref{tvm})
and using  Eq.~(\ref{p0m})
shows that
\begin{equation}P^-_{MF}= P_0^-(N).
\end{equation}  
  The  ground state
eigenvector of this operator  is a Slater
determinant denoted as $\mid\phi\rangle$:  
\begin{equation} P^-_{MF}\mid\phi\rangle
  =P_0^-(N)\mid\phi\rangle
= m_0\mid\phi\rangle.\label{phi}
\end{equation}

We shall use both the Hartree-Fock and Bruckner Hartree Fock approximations
to obtain expressions for  $\widehat{U}$.
For now we pursue the question:
Given a $\widehat{U}$, how do we proceed? 
The first step is to expand the field operator $\psi$ in terms of the
eigenfunctions of $i\partial^-$ in the light front Dirac equation (\ref{mffe})
The nucleon field operator is constructed as follows:
\begin{eqnarray}
\psi (x) = \int {d^2k_\perp dk^+  \theta (k^+)\over (2\pi)^{3/2} \sqrt{2k^+}}
\sum_\lambda u (k, \lambda)
e^{-ik\cdot x},
\end{eqnarray}
where $k\cdot x={1\over 2}\left(k^-x^++k^+x^-\right)-\bbox{k}_\perp\cdot
\bbox{x}_\perp$.
We keep only the nucleon part of $\psi(x)$ as the anti-nucleon degrees of
freedom are not needed here. These spinors    
$u(k,\lambda)$ are the eigenfunctions
of Eq.~(\ref{mffe}), with normalization
$\bar{u}(k, \lambda)\gamma^+u(k, \lambda) =2k^+$.
For the present treatment of the  translationally invariant
infinite nuclear matter
system, the mean-field potentials $U_S$ and $U_V^-$ are independent
of the spatial position $\;x_\perp,x^-$. The eigenvalues of Eq.~(\ref{mffe})
are given by\cite{gam97a,gam97b}
\begin{eqnarray}
k^-=U_V^- + {\bbox{k}_\perp^2 + (M+U_S)^{2} \over k^+}, \label{sp}
\end{eqnarray}
in which  $U_S$ and $U_V$ depend upon $\bbox{k}_\perp$ and $k^+$. 
%

The next step is to better define the Slater determinant $\mid\phi\rangle$.
The occupied states are to 
 fill up a Fermi sea, which is usually defined in terms of a Fermi momentum,
  $k_F$, that 
is the magnitude of a three vector. This three vector is defined\cite{gs} as:
\begin{equation}
k^+=\sqrt{(M+U_S)^2+{\bbox k}\cdot{\bbox k}}+k^3,\label{kplus}
\end{equation}   
which implicitly defines $k^3$.  Using Eq.~(\ref{kplus}) allows one to
maintain the equivalence between energies computed in the light front
and equal time formulations of scalar field theories\cite{bg} and to 
 restore manifest rotational
invariance in light-front QED\cite{mp}.

The computation  of the energy and plus 
momentum distribution proceeds from taking the appropriate expectation
values of the energy 
momentum tensor $T^{\mu\nu}_{\cal V}$ . 
\begin{equation}
P^\mu_{\cal V}
={1\over 2}\int d^2 x_\perp dx^- \langle\phi\mid T^{+\mu}_{\cal V}
\mid\phi\rangle.\label{pmu}
\end{equation}
We are concerned with the light front energy $P^-$ and momentum $P^+$.
The relevant components of 
$T^{\mu\nu}_{\cal V}$ are presented in Eqs.~(\ref{tvm}) and (\ref{tvp}).
 Taking the nuclear matter expectation
value of $T^{+-}_{\cal V}$ and $T^{++}_{\cal V}$
and performing the spatial integral of 
Eq. (\ref{pmu}) leads to the  result 
\begin{eqnarray}
{P^{-}_{\cal V}\over \Omega}&=&
{4\over (2\pi)^3}\int_F d^2k_\perp dk^+\left\{ {\bbox{k}_\perp^2+ 
(M+U_S)^2\over k^+}-\;2\; 
{1\over 2}\sum_\lambda\;
{\bar{u}(k,\lambda)\over\sqrt{2k^+}}U_S{u(k,\lambda)\over\sqrt{2k^+}}\right\}\;
+\langle\phi\mid{K}\mid\phi\rangle
\label{pminus}\\
{P^{+}_{\cal V}\over \Omega}&=&
{4\over (2\pi)^3}\int_F d^2k_\perp dk^+ k^+,\label{pplus}
\end {eqnarray}
where $\Omega$ is the volume of the system
$\Omega\equiv {1\over 2} \int d^2x_\perp dx^-$.
The subscript F denotes that $\mid\vec k\mid<k_F$   with $k^3$ defined
by the relation (\ref{kplus}). The integral
involving $\bar{u}(k,\lambda)U_Su(k,\lambda)$ may also be expressed as:
\begin{equation}
\langle\phi\mid U_S\mid\phi\rangle
\equiv {4\over (2\pi)^3}\int_F d^2k_\perp \;dk^+
\;{1\over 2}\sum_\lambda{\bar{u}(k,\lambda)\over\sqrt{2k^+}}U_S
{u(k,\lambda)\over\sqrt{2k^+}}.
\end{equation}

Equations (\ref{pminus}) and (\ref{pplus}) along with the expression
for $k^+$, (\ref{kplus}) allow an evaluation of $P^-$ and $P^+$.
This shall be done by obtaining 
the mass $M_0$ of the A-nucleon system as
$M_0={1\over 2}(P^+_{\cal V}+P^-_{\cal V})$ and minimizing $M_0$ per nucleon.
For a nuclear system at rest, the eigenvalues of
the plus and minus momentum operators must be the same, so it should
be legitimate to use the average value. In practise, approximations are made.
However, the 
 procedure of minimizing the average ${1\over 2}(P^+_{\cal V}+P^-_{\cal V})$
 is the same as
minimizing the expectation value of $ P^{-}_{\cal V}$
subject to the constraint that the expectation value of $ P^{+}_{\cal V}$
is equal to the value of $ P^{-}_{\cal V}$
\cite{gam97b}.
Summing equations (\ref{pminus}) and (\ref{pplus}) and dividing by a factor
of 2 leads to 
\begin{eqnarray}
{M_0\over\Omega}=
{4\over (2\pi)^3}{1\over2}\int_F d^2k_\perp dk^+ \left({\bbox{k}_\perp^2+ 
(M+U_S)^2\over k^+}+k^+\right)-\langle\phi\mid U_S\mid\phi
\rangle+\langle\phi\mid{K\over 2}\mid\phi\rangle.
\label{Ei}
\end {eqnarray}
Then replace the integration over $k^+$ by one over $k^3$ using
Eq.~(\ref{kplus}) and the definition
\begin{equation}
E^*_{\bbox{k}}\equiv \sqrt{\bbox{k}\cdot\bbox{k}+
(M+U_S)^2}, \label{estar}
\end{equation}
so that  Eq.~(\ref{Ei}) takes the form:
\begin{equation}
{M_0\over \Omega}= 
{4\over (2\pi)^3}\int d^3k\theta(k_F-k)\;E^*_{\bbox{k}}
-\langle\phi\mid U_S\mid\phi\rangle+\langle\phi\mid{K\over 2}\mid\phi\rangle, 
\label{E}
\end{equation}
with
\begin{eqnarray}\langle\phi\mid U_S\mid\phi\rangle=
{4\over (2\pi)^3}\int d^3k\theta(k_F-k)
{1\over2}\sum_\lambda{\bar{u}(k,\lambda)\over\sqrt{2E^*_{\bbox{k}}}}U_S
{u(k,\lambda)\over\sqrt{2E^*_{\bbox{k}}}}\nonumber\\
={4\over (2\pi)^3}\int d^3k\theta(k_F-k){2(M+U_S)\over 2E^*_{\bbox{k}}}
  U_S.
\end{eqnarray}

One obtains a formalism that looks more conventional by using a discrete
representation of the single nucleon states. We define  a set of spinors
$\mid\alpha\rangle$, with $\alpha $ representing the quantum numbers
$\bbox{k}$ and  $\lambda$ such that
\begin{eqnarray}
  \langle x\mid\alpha\rangle&\equiv& {e^{-ik\cdot x}\over\sqrt{ \Omega}}
  u(k,\lambda)\sqrt{1\over 2E^*_{\bbox{k}}}\nonumber\\
  \langle\bar{\alpha}\mid&\equiv&\langle{\alpha}\mid\gamma^0\label{alphas}
  \nonumber\\ 1&=&{1\over 2}\int d^2x_\perp
  dx^- \langle \alpha\mid x\rangle\langle
  x\mid\alpha\rangle. \label{newnorm}
 \end{eqnarray} 
The difference between $\langle x\mid\alpha\rangle$ and a usual equal time
(ET)
spinor, $\langle x\mid\alpha\rangle_{ET}$,
of the same quantum numbers,  energy $\epsilon(k)=E^*_{\bbox{k}} 
+U_V^0$, and
normalization 
can be determined by considering the
phase factor, using Eq.~(\ref{sp}):
\begin{eqnarray}%
  k\cdot x & =&
  {k^-x^+\over 2}+ {k^+x^-\over 2}-\bbox{k}_\perp\cdot\bbox{x}_\perp\nonumber
  \\
   & =& \left(U^-_V+{\bbox{k}_\perp^2+(M+U_s)^2\over k^+}\right)
{(t+z)\over 2}
 + k^+{(t-z)\over 2}-\bbox{k}_\perp\cdot\bbox{x}_\perp\nonumber\\
   & =& \epsilon(k)t -\vec{k}\cdot\vec{r} -i{U_V^0x^-\over 2}. \label{kdx}
 \end{eqnarray}
 The last factor is the consequence of using the barred form of the vector
 potential according to Eq.~(\ref{transmf}). The only difference between the
 light front spinors and those of the equal time form is due to this
 phase factor. The consequences of this phase factor for computations of the
 light front momentum density are that nucleons carry only 65\% of the
 nuclear plus momentum in the mean field calculation\cite{gam97a,gam97b}.
 A similar result is to be found below. Furthermore this phase factor has the
 desirable feature of suppressing the number of nuclear anti-nucleons
 \cite{bbm98}.

 For now, we consider the effects
 of using light front spinors on the calculation of the energy.
 We express the energy $M_0$ of Eq.~(\ref{E}) in terms of the spinors
 $\mid\alpha\rangle $ of  Eq.~(\ref{alphas}) as: 
  \begin{equation}
    M_0=\sum_{\alpha<F} E_\alpha-
   \sum_{\alpha<F}\langle\bar{\alpha}\mid U_S\mid\alpha\rangle
   +{1\over 2}\sum_{\alpha,\beta<F}
   \langle\bar{\alpha}\bar{\beta}\mid V
   \mid\alpha\beta\rangle_a, 
 \end{equation}
 in which $\mid\alpha\beta\rangle_a\equiv\mid\alpha\beta\rangle-
 \mid\beta\alpha\rangle$. In this notation
  \begin{equation}
   E_\alpha=
   \epsilon_\alpha-\langle\bar{\alpha}\mid\gamma^0U_V^0\mid\alpha\rangle,
  \end{equation}
  so that we obtain
\begin{equation}M_0=\sum_{\alpha<F} \epsilon_\alpha-
   \sum_{\alpha<F}\langle\bar{\alpha}\mid U_S+\gamma^0U_V^0\mid\alpha\rangle
   +{1\over 2}\sum_{\alpha,\beta<F}
   \langle\bar{\alpha}\bar{\beta}\mid V
   \mid\alpha\beta\rangle_a. \label{ea}
 \end{equation}
 This result gives the value of the nuclear energy in terms of the
 eigenvalues of the light front Dirac equation and in terms of light front
 spinors. The term $K/2$ has been replaced by $V$ according to Eq.~(\ref{kdef})
 and using the normalization of Eq.~(\ref{alphas}).

 It is necessary
 however, to consider the effects of the phase factor 
  $-i{U_V^0x^-\over 2}$ of Eq.~(\ref{kdx}) which accounts for the difference
  between light front and equal time spinors.
  This factor and its complex
  conjugate multiply  to unity in the calculation of
  the matrix elements $\langle\bar{\alpha}\mid
  U_S+\gamma^0U_V^0\mid\alpha\rangle$ and in the matrix element
  $ \langle\bar{\alpha}\bar{\beta}\mid V\mid\alpha,\beta\rangle$.
  We need to consider also the matrix element
  $ \langle\bar{\alpha}\bar{\beta}\mid V\mid\beta\alpha\rangle$ for which
  the effects
  of the phase factor do not automatically cancel.
  In principle,
  $U_V^0$ is a function of the momentum denoted by the quantum numbers
  $\alpha,\beta$. In practise, this dependence is weak and  can be ignored
  in calculations of the energy, provided one evaluates the potential at a
  reasonably chosen average value. Furthermore, in evaluating the matrix
  element  $ \langle\bar{\alpha}\bar{\beta}\mid V\mid\beta\alpha\rangle$
  both of the states $\beta\alpha$ are below the Fermi sea and have a
  momentum separated by an amount small compared to the scale of the momentum
  dependence. Thus the phase factor does not enter  in present calculations
  of the energy
  (but does in the evaluation of the plus-momentum distribution).
  Our calculations of the energy in the light front and equal
  time formulations yield the same results. However calculations of the
  plus-momentum distributions  can only be done using
  the light front formalism.

  Let us determine the mean field, and the corresponding value of $M_0$
  in terms of $V$.
 The light front Hartree-Fock (HF) approximation is defined by taking
the mean field to be calculated from the average potential according to
\begin{eqnarray}
 \langle\bar{\alpha}\mid (U_S+\gamma^0U_V^0)^{HF}\mid\alpha\rangle&\equiv&
U(\alpha)= \sum_{\beta<F}
   \langle\bar{\alpha}\bar{\beta}\mid V
   \mid\alpha\beta\rangle_a. 
\end{eqnarray}
Summing over the occupied orbitals gives
\begin{eqnarray}  \sum_{\alpha<F}  U(\alpha) 
&=& 2\langle\phi\mid V
\mid\phi\rangle.
\end{eqnarray}
In this case,  the use of the above equation in the expression (\ref{ea})
for $M_0$,  leads to the HF approximation for the nuclear energy:
\begin{equation}M_0^{HF}=\sum_{\alpha<F} \epsilon_\alpha
   -{1\over 2}\sum_{\alpha,\beta<F}
   \langle\bar{\alpha}\bar{\beta}\mid V
   \mid\alpha\beta\rangle_a, 
 \end{equation}
 which
 has the same form as the expression for the energy
 in the usual equal time Hartree Fock expression.

We shall also need to obtain $m_0$, the eigenvalue of $P^-_{MF}$. This
expression will be used in obtaining
the Bruckner Hartree-Fock approximation.
  According to
  Eqs.~(\ref{tvm}) and (\ref{phi}), the difference between $M_0$ and $m_0$ is
  the expectation value of the potential $V$.  Thus  
\begin{equation}m_0=\sum_{\alpha<F} \epsilon_\alpha-
   \sum_{\alpha<F}\langle\bar{\alpha}\mid U_S+\gamma^0U_V^0\mid\alpha\rangle
   \label{m0}
 \end{equation}

\subsection{Light Front Brueckner  Hartree-Fock Approximation}
The interaction $K$ between two nucleons is  strong and  
the scattering amplitude is obtained, as discussed in Section 3,
by solving the Weinberg
equation---the 
light front version of the Lippmann-Schwinger equation. Thus we need to go
beyond the Hartree-Fock approximation. This shall be accomplished by
treating the interaction between two nucleons to all orders in $K$. 

The idea is that we wish to find the Slater determinant
$\mid\phi\rangle$, recall Eq.~(\ref{phi}),
that leads to the best approximation for the energy $M_0$ of
the full nucleonic wave function
 $\mid \Phi\rangle$, recall Eq.~(\ref{big}).

 Both of the states 
$\mid\phi\rangle$ and  $\mid \Phi\rangle$ 
 are eigenstates of a P-minus operator, and both are eigenstates
of the operator $P_0^+(N)$. 
   We shall use standard techniques to derive a perturbation theory in
   $K $ 
   to obtain an expression for  the state $\mid\Phi\rangle$
   in terms of  $\mid\phi\rangle$.
Thus we write  
 \begin{eqnarray}
\mid \Phi \rangle &=& \mid \phi \rangle  + \Lambda \mid \Phi \rangle 
 \label{decomp}
\end{eqnarray}
with
\begin{eqnarray}
\Lambda &\equiv& 1 - \mid \phi \rangle \langle \phi \mid .
\end{eqnarray}
Then use Eq.~(\ref{decomp}) in Eq.~(\ref{big}) and multiply the result
on the
left by $\Lambda $ to obtain 
\begin{equation} 
\Lambda \mid \Phi \rangle  = 
{1\over M_0 - \Lambda( P_0^-(N)+K)\Lambda} 
\Lambda K
\mid \phi \rangle ,\end{equation} 
so  that
\begin{eqnarray}
\mid \Phi \rangle  &=& \mid \phi \rangle 
 + {1\over M_0 - \Lambda ( P_0^-(N)+K) \Lambda} 
\Lambda K 
\mid \phi \rangle . \label{Phi}
\end{eqnarray}
We can obtain a useful expression for $M_0$ by acting with the operator
$ \langle\phi\mid \left(P_0^-(N)+K\right)$ 
on the left of  Eq.~(\ref{Phi}) and using
the result  $\langle \phi \mid \Phi \rangle  = 1,$ which follows from
Eq.~(\ref{Phi}). 
Then we find
\begin{eqnarray}
M_0 = \langle \phi \mid P^-_0 +K\mid \phi \rangle 
+ \langle\phi \mid K
\Lambda {1\over M_0-\Lambda ( P_0^-(N)+K)\Lambda}
\Lambda K
\mid \phi \rangle .\label{M0}
\end{eqnarray}
Using Eq.~
(\ref{phi})
in Eq.~(\ref{M0}) leads to 
\begin{eqnarray}
M_0 - m_0 = \langle \phi \mid K\mid \phi\rangle 
+ \langle \phi \mid K
\Lambda {1 \over M_0- \Lambda ( P_0^-(N)+K)\Lambda}
\Lambda K 
\mid \phi \rangle  , \label{mass}
\end{eqnarray}
which can be re-stated as
\begin{eqnarray}
M_0 - m_0 = \langle \phi \mid X\mid \phi\rangle,\label{mmm}
\end{eqnarray}
where
\begin{eqnarray}
  X=K + K\Lambda{1\over M_0-\Lambda P_0^-(N)\Lambda}\Lambda X.
\end{eqnarray}
The operator $X$ is a many-body operator acting 
on all nucleons via the iterations
of the two-nucleon  interaction $K$. We shall make the independent pair
approximation of including only pair-wise interactions. Thus we approximate 
\begin{eqnarray}\langle \phi \mid X\mid \phi\rangle\approx
\langle \phi\mid{1\over2}\sum_{i,j} \Gamma_{i,j}(P^-_{ij})\mid \phi\rangle
\equiv\langle \phi \mid  \Gamma\mid \phi\rangle, \label{pgp}
\end{eqnarray}
where $\Gamma_{i,j}$ is a two-nucleon operator which is a solution of the
integral equation
\begin{eqnarray}
\Gamma_{i,j}(P^-_{ij}) &=& K_{ij} + K_{ij} \,\,{\Lambda\over
 P^-_{ij}-\Lambda P_0^-(N)\Lambda} \,\, \Gamma_{i,j}(P^-_{ij}).
\end{eqnarray}
The notation $i,j$ refers to a pair of particles. The relevant matrix element
is expressed using 
 the  eigenstates of Eq.~(\ref{mffe})
as
\begin{equation} 
\langle3,4|\Gamma(P^-_{1,2})|1,2\rangle
=\langle 3,4|
K|1,2\rangle+
\sum_{\lambda_5,\lambda_6}\int  \langle 3,4|
K|5,6\rangle 
{2{M^*}^2\over p_5^+p_6^+}
{d^2p_{5\perp}dp^+_5\; Q\over P^-_{1,2}-(p_5^-+p_6^-)+i\epsilon}
\langle5,6|\Gamma|1,2\rangle, \label{bweinberg}
\end{equation}
in which we define
\begin{equation}
  M^*\equiv M+U_S. \label{mstar}
  \end{equation}
The operator $Q$, to be specified below,  is the two-body version of
$\Lambda$ and
projects the momenta $p_5$ and $p_6$ above the
Fermi sea. 
The factor $ {{M^*}^2\delta^{(2,+)}(P_i-P_f)\over \sqrt{k_1^+k_2^+k_3^+k_4^+}}$
appears in each of the terms of Eq.~(\ref{bweinberg}), so it is worthwhile
to define a Bruckner $G$-Matrix $G$ using
\begin{equation}
  \Gamma\equiv 2
  G {{M^*}^2\delta^{(2,+)}(P_i-P_f)\over \sqrt{k_1^+k_2^+k_3^+k_4^+}}.
\end{equation}

To follow the steps of Sect. III in converting Eq.~(\ref{bweinberg}) into
one of a more familiar  form, in which rotational invariance is manifest,
one needs to know the values of
\begin{equation}P_{1,2}^-=p_1^-+p_2^-,
  \end{equation}
  which for the case of
relevance here in computing the nuclear expectation value in the independent
pair approximation, is the
same as $p_5^-+p_6^-$.
The single-particle minus-momentum 
eigenvalues are  given according to Eq.~(\ref{sp}) as 
\begin{eqnarray}
p^-_i
&=& {{\bbox{k}_i}_\perp^2 +(M+U_S)^2\over k^+_i} +U_V \label{pim}
\end{eqnarray}
Our approximation is that $U_V$ is independent of orbital $i$. Thus this
potential cancels out in computing the difference $ P_i^--(p_5^-+p_6^-)$
and the energy denominator is as in the free space considerations of Sect.~III,
except that the mass of the nucleon is replaced by $M+U_S$. Thus the previous
derivation of an equivalent three-dimensional integral equation that is
manifestly covariant and rotationally invariant proceeds as before.

One expresses  the plus-momentum   variable in terms of a light-front 
momentum  fraction
$\alpha$ of Eqs.~(\ref{alpha}) and (\ref{alpha2}) so that one obtains
\begin{equation} 
\langle3,4|
G|1,2\rangle
=\langle 3,4|
V|1,2\rangle+
\int\sum_{\lambda_5,\lambda_6} \langle 3,4|
V|5,6\rangle  
{2{M^*}^2\over \alpha(1-\alpha)}
{d^2p_\perp d\alpha\;Q\over P_i^2-{p_{\perp}^2+{M^*}\;^2\over\alpha(1-\alpha)}
}
\langle5,6|G
|1,2\rangle, \label{519}
\end{equation}
where $P_i^2$ is square of the total initial  four-momentum, computed using
$M+U_S$ for the nucleon mass.

Equation~(\ref{519}) can
in turn be re-expressed as a medium-modified
Blankenbecler-Sugar (BbS) equation
\cite{BbS} 
by  using the  medium-modified version of the
variable transformation\cite{Te 76}:
\begin{equation}
\alpha={E^*_{\bbox{p}}+p^3\over 2E^*_{\bbox{p}}}, \label{balpha}
\end{equation}
with $E^*_{\bbox{p}}$ 
given in Eq.~(\ref{estar}). The result is:
\begin{equation}
\langle3,4|G
|1,2\rangle
=\langle3,4|V
|1,2\rangle+\int
\sum_{\lambda_5,\lambda_6} \langle 3,4|V
|5,6\rangle
{{M^*}^2\over E^*_{\bbox{p}}}
{d^3p\; Q \over p_i^2-p^2}
\langle5,6|G
|1,2\rangle, \label{bbsbs}
\end{equation}
which is the desired equation 
(with Dirac spinors normalized as in Sect.~III).

 The Brueckner light front Hartree-Fock (BHF)
 approximation is defined by taking
the mean field to be calculated from the average G-matrix according to
\begin{eqnarray}
   U(\alpha)=
 \langle\bar{\alpha}\mid (U_S+\gamma^0U_V^0)^{BHF}\mid\alpha\rangle&=&
 \sum_{\beta<F}
   \langle\bar{\alpha}\bar{\beta}\mid G
   \mid\alpha\beta\rangle_a.
\end{eqnarray}
where Dirac spinors are normalized as in Eq.~(\ref{alphas}).
The sum over occupied orbitals gives
\begin{eqnarray}\sum_{\alpha<F} U(\alpha)
  &=& 2\langle\phi\mid G
\mid\phi\rangle.
\label{ubhf}
\end{eqnarray}
We use  this BHF mean-field to determine the value of $m_0$ via 
Eq.~(\ref{m0}). Then the use of Eq.~(\ref{pgp}) in Eq.~(\ref{mmm})
determines the value of $M_0$ as the eigenvalue of $P^-_{\cal V}$.
But $M_0$ is also the eigenvalue (or in this case the expectation value)
of  $P^+_{\cal V}$. The  minimization  of $P^-_{\cal V}$ subject to the
constraint that the expectation value of $P^+_{\cal V}$ is the value
of $P^-_{\cal V}$ leads to the BHF version of Eqs.~(\ref{pminus}) and
(\ref{pplus}):
\begin{eqnarray}
{P^{-}_{\cal V}\over \Omega}&=&
{4\over (2\pi)^3}\int_F d^2k_\perp dk^+\left\{ {\bbox{k}_\perp^2+ 
(M+U_S)^2\over k^+}-2\;
{1\over 2}\sum_\lambda\;
{\bar{u}(k,\lambda)\over \sqrt{2 k^+}}U_S
{u(k,\lambda)\over\sqrt{2 k^+}}\right\}\;
+\langle\phi\mid{\Gamma}\mid\phi\rangle
\label{pbminus}\\
{P^{+}_{\cal V}\over \Omega}&=&
{4\over (2\pi)^3}\int_F d^2k_\perp dk^+ k^+.\label{pbplus}
\end {eqnarray}
Note that the quantity $k^+$ is defined in Eq.~(\ref{kplus}). 

Taking the average of  equations (\ref{pbminus})  and~(\ref{pbplus}), and
using the basis of Eq.~(\ref{alphas}) leads to our result for the BHF version
of the nuclear mass:
\begin{equation}M_0=\sum_{\alpha<F} \epsilon_\alpha
   -{1\over 2}\sum_{\alpha,\beta<F}
   \langle\bar{\alpha}\bar{\beta}\mid G
   \mid\alpha\beta\rangle_a, \label{main}
 \end{equation}
This is equivalent to the usual expression of
Ref.~\cite{Mac89,rm1}, cf.\ Eqs.~(\ref{E/A0}) and (\ref{E/A}) below.

\section{Light-Front Brueckner Theory of Nuclear Matter}

\subsection{Summary}
The formalism of the previous section can be summarized using the notation
of Ref.~\cite{Mac89,rm1}. In that work,  
single-nucleon motion in nuclear matter is described by the Dirac
equation
\begin{equation}
(\not\!k-M-U)u^*({\bbox k},s)=0
\label{Dirac}
\end{equation}
or in Hamiltonian form
\begin{equation}
(\mbox{\boldmath $\alpha$} \cdot {\bbox k} + \beta M + \beta U)
u^*({\bbox k},s) = \epsilon_{k} u^*({\bbox k},s)
\end{equation}
with
\begin{equation}
U=U_{S} +\gamma^{0}U^0_{V} \; ;
\end{equation}
where we use the usual notation~\cite{Bj}, $\beta\equiv \gamma^0$
and $\bbox{\alpha}\equiv \gamma^0 \bbox{\gamma}$.
The solution of Eq.~(\ref{Dirac}) is
\begin{equation} 
u^*({\bbox k},s)=
\sqrt{\frac{E^*_{\bbox k}+M^*}{2E^*_{\bbox k}}}
  \left( \begin{array}{c} 1\\ 
\frac{\bbox{\sigma} \cdot \bbox{k}}
{E^*_{\bbox k}+M^*} \end{array} \right) \chi_{s}
\label{InMedSpinor}
\end{equation}
with $M^*$ defined in Eq.~(\ref{mstar}).
 $E^*_{\bbox k}$ given by Eq.~(\ref{estar}),
and $\chi_{s}$ a Pauli spinor.
The normalization is
 $\bar{u^*}({\bbox k},s)u^*({\bbox k},s)={E^*_{\bbox k}\over M^*}$
and
 ${u^{*\dagger}}({\bbox k},s)u^*({\bbox k},s)=1$, as in Eq.~(\ref{newnorm}).
Notice that the in-medium Dirac spinor Eq.~(\ref{InMedSpinor}) 
is obtained from the free Dirac spinor
by simply replacing $M$ by $M^*$. The difference between the equal time spinors
used in Refs.~\cite{Mac89,rm1} 
and light front spinors is the phase factor discussed above in Sect. IV.C.
As noted,
this factor cancels out of the matrix elements when using the approximation
that the vector potential is independent of momentum.

The nuclear matter $G$-matrix is the solution of the integral
equation,
(\ref{bbsbs})  without the explicit spin indices,
\begin{equation}
G({\bbox k'}, \bbox{k};|{\bbox P}|,k_F)=V^*({\bbox k'},{\bbox k})+
\int \frac{d^{3}p}{(2\pi)^{3}}
 V^*({\bbox k'}, {\bbox p}) \frac{{M^*}^{2}}
{E^*_{\frac{1}{2}{\bbox P}+{\bbox p}}}
 \frac{\bar{Q}(|{\bbox p}|,|{\bbox P}|,k_F)}
{{\bbox k}^2-{\bbox p}^2+i\epsilon}
G({\bbox p},\bbox{k};|{\bbox P}|,k_F) \; ,
\label{GEQ}
\end{equation}
where
${\bbox P}$ is the total three-momentum of the two nucleons
in the nuclear matter rest frame;
and ${\bbox k}$, $\bbox{p}$ and ${\bbox k'}$ are
the initial, intermediate and final relative momenta, respectively,
of the two nucleons interacting in nuclear matter. Note that the third
components of the various momenta are obtained by using the appropriate
versions of Eq.~(\ref{kplus}). 
$k_F$ denotes the magnitude of the Fermi momentum 
corresponding to the nuclear matter density
under consideration.
The Pauli operator $Q$, for which we use the angle average
$\bar{Q}$~\cite{HT70},
projects onto unoccupied states. 
In the derivation, we have used the angle averages
$(\:\frac12{\bbox P}\pm{\bbox k}\:)^2\approx\frac14 {\bbox P}^2 + {\bbox k}^2$
and 
$(\:\frac12{\bbox P}\pm{\bbox p}\:)^2
\approx\frac14 {\bbox P}^2 + {\bbox p}^2$.  
The latter implies
$E^*_{\frac{1}{2}{\bbox P}+{\bbox p}}
\approx\sqrt{{M^*}^2+\frac14 {\bbox P}^2 + {\bbox p}^2}$.  

The essential difference with standard Brueckner theory is the use
of the potential
 $V^*$ in Eq.~(\ref{GEQ}). As indicated by the asterix,
the OBE potential  
of Eqs.~(\ref{VOBEpv})-(\ref{VOBEv}) is evaluated
by using the in-medium spinors Eq.~(\ref{InMedSpinor}) 
instead of the free ones.

It is necessary to discuss a
technical point concerning the retardation effects in the form factors
Eq.~(\ref{Cutoff1})
that enter in evaluating $V^*$.
Consider the matrix element of $V^*$ for which our formalism says to use
\begin{equation}q^0=E^*_\alpha-E^*_\beta=(E^*_\alpha+U_V(\alpha))-(E^*_\beta
  +U_V(\alpha)).
  \end{equation}
We have consistently ignored the state dependence of $U_V$. This is a good
approximation for energy differences governed by the Fermi energy. However,
the relevant energy scale in the form factors are the parameters $\Lambda$ of
Table 1, which are on the order of a GeV or more.
Thus for these terms it would be better to use
\begin{equation}q^0=(E^*_\alpha+U_V(\alpha))-(E^*_\beta
  +U_V(\beta)). \label{try}
  \end{equation} The quantity $E^*_\alpha+U_V(\alpha)$ for occupied orbitals
  $\alpha $ is close to the nucleon mass. For high energy orbitals $\beta$, 
$U(\beta)$ is small. Thus the use of Eq.~(\ref{try}) 
 is
 numerically very similar to ignoring the effects of the medium modifications
in evaluating the retardation effects in the form factors, which is what we do.

 The single-particle potential
for nucleons in nuclear matter is
\begin{equation}
U(\alpha)=
\langle \bar{\alpha}| U | \alpha \rangle =
 \langle \bar{\alpha}|U_{S}+\gamma^{0}U^0_{V}|\alpha\rangle
=
\frac{M^*}{E^*_\alpha} U_{S} +U^0_{V} \; ;
\label{U1}
\end{equation}
which is calculated from the $G$-matrix by
\begin{equation}
U(\alpha)=Re\sum_{\beta<{F}}
 \langle \bar{\alpha}\bar{\beta}|G|\alpha\beta\rangle_a
\label{U2}
\end{equation}
where $\alpha$ denotes a state below or above the Fermi surface 
(continuous choice). 

Using the notation of the previous section, we define
`the energy per nucleon in nuclear matter' by 
\begin{equation}
\frac{{\cal E}}{A}=\frac{M_0}{A} - M \; ,
\label{E/A0}
\end{equation}
which can also be written as
\begin{equation}
\frac{{\cal E}}{A}=\frac{1}{A}
\sum_{\alpha<{F}}
\langle \bar{\alpha} | \bbox{\gamma} \cdot {\bbox k}_{\alpha}
+ M | \alpha \rangle 
+ \frac{1}{2A} \sum_{\alpha,\beta<{F}}
\langle \bar{\alpha}\bar{\beta}|G|\alpha\beta\rangle_a -M
\; .
\label{E/A}
\end{equation}
Note that this equation depends on the Fermi momentum $k_F$ and, thus,
on the density of nuclear matter.

It is useful to have the following summary of formulae
concerning the single particle energy $\epsilon_\alpha$:
\begin{eqnarray}
\epsilon_\alpha
 & = & \langle \bar{\alpha} | \bbox{\gamma} \cdot {\bbox k}_{\alpha}
+ M + U | \alpha \rangle  \\
 & = & \langle \bar{\alpha} | \bbox{\gamma} \cdot {\bbox k}_{\alpha}
+ M | \alpha \rangle  + U(\alpha)  \\
 & = & \frac{MM^*+\bbox{k}^2_\alpha}{E^*_\alpha}
                    +  \frac{M^*}{E^*_\alpha} U_{S} +U^0_{V} \\
 & = & E^*_\alpha + U^0_V \; ,
\end{eqnarray}
with
$E^*_\alpha\equiv \sqrt{M^{*2}+\bbox{k}^2_\alpha}$
and $M^*=M+U_S$.

The calculation of the nuclear matter $G$ matrix involves a self-consistency,
since the solution of Eq.~(\ref{GEQ}) for $G$ requires knowledge
of $M^*$ which, in turn, is determined from $G$ via 
Eqs.~(\ref{U1}) and (\ref{U2}). 
In practise, one starts out with an
educated guess for $M^*$, solves Eq.~(\ref{GEQ}) for $G$
and uses this $G$ to calculate a new $M^*$ from
Eqs.~(\ref{U1}) and (\ref{U2}). 
The procedure is then repeated starting with the new $M^*$.
This is reiterated until 
the calculated $M^*$ reproduces accurately
the starting $M^*$.

\subsection{Results}
The formalism of the previous section is used to  calculate
the energy per nucleon in nuclear matter as a function of density,
Eq.~(\ref{E/A}). Our result is plotted in Fig.~2 by the solid line.
The curve saturates at ${\cal E}/A = -14.71$ MeV and $k_F = 1.37$
fm$^{-1}$, and predicts an incompressibility of $K=180$ MeV at the minimum.
These predictions agree well with the
empirical values 
${\cal E}/A=-16\pm 1$ MeV,
$k_F=1.35\pm 0.05$ fm$^{-1}$, 
and $K=210\pm 30$ MeV~\cite{Bla80}.

To get a better idea of the quality of our predictions, it is useful to
compare with the results from alternative relativistic approaches.
Brockmann and Machleidt~\cite{rm1} predict
${\cal E}/A=-13.6$ MeV,
$k_F=1.37$ fm$^{-1}$, 
and $K=250$ MeV at saturation, using the equal-time formalism
and their `Potential B'.
The greatest difference occurs for the incompressibility
which is predicted smaller by the LF Brueckner theory implying a softer
equation of state. This can be partially attributed to the 
medium effect that comes
from the meson propagators in the LF approach and that is absent in the 
equal time (ET) approach. Recall that in the LF formalism the momentum
transfer between two nucleons exchanging a meson is
\begin{equation}
q = 
 (q_0,\bbox{q}) = 
(E'-E,{\bbox k'}-{\bbox k}) \; ,
\label{TransLF}
\end{equation}
where $E$ and $E'$ are nucleon on-mass-shell energies
[more explanations can be found below Eq~(\ref{VOBEv})],
implying the meson propagators
\begin{equation}
\frac{i}{q^2-m_\alpha^2}=
\frac{i}{(E'-E)^2-({\bbox k'}-{\bbox k})^2-m_\alpha^2} \; ,
\label{PropLF}
\end{equation}
while in the ET formalism no energy is transfered, thus,
\begin{equation}
q = (0,\bbox{q}) = 
(0,{\bbox k'}-{\bbox k}) \; ,
\label{TransET}
\end{equation}
and the propagators are
\begin{equation}
\frac{i}{-{\bbox q}^2-m_\alpha^2}=
\frac{i}{-({\bbox k'}-{\bbox k})^2-m_\alpha^2} \; .
\label{PropET}
\end{equation}
In nuclear matter, the free-space LF meson propagators Eq.~(\ref{PropLF})
are replaced by
\begin{equation}
\frac{i}{(E'^*-E^*)^2-({\bbox k'}-{\bbox k})^2-m_\alpha^2} \; ,
\label{PropLFinmed}
\end{equation}
while the ET propagators undergo no changes.
The medium effect on the LF meson propagators enhances them
off-shell which leads to more binding energy.
This is demonstrated
in Fig.~2 where the difference between the dotted and solid curve 
is generated by the medium effect on the meson propagators.

There is another difference that arises from a technical issue in the
solution of the transcendental equation for the
G-matrix. We obtain new values of the mean fields $M^*(k_F)=M_N+U_S=
718 $ MeV and $U_V=165$ MeV. The mean field potentials  obtained here
from the G-matrix 
are considerably 
smaller than those of mean field theory in which the potential is used.
We discuss the implications for nuclear deep inelastic scattering
 in Sec.~VII.

The most important medium effect in
relativistic approaches to nuclear matter comes from the
use of in-medium Dirac spinors representing the nucleons
in nuclear matter (`Dirac effect').  This effect (and the medium effect on
meson propagators) is absent in the conventional
(nonrelativistic) Brueckner calculation which yields the dashed
curve in Fig.~2. 
Characteristic for all predictions using conventional Brueckner theory is
that the saturation density is predicted too high and, thus, they all fail
to explain nuclear saturation correctly.

The effect that is generated by the in-medium Dirac
spinors is strongly density dependent (due to the density
dependence of $M^*$) shifting the saturation
curve towards lower densities such that nuclear saturation
is predicted at the correct density (solid and dotted curves in Fig.~2).

The effect from the in-medium Dirac spinors is, of course, largest
for $\sigma$ and $\omega$ exchange for which the LF and ET formalisms
predict essentially the same. However, $\pi$ and $\rho$
do also contribute to the medium effect
and, here, we have differences between LF and ET.
The general underlying reason for this difference is that
for derivative coupling, implying a momentum dependence of the
meson-nucleon vertex, the difference in the momentum transfer 
(= meson momentum) between LF and ET [Eqs.~(\ref{TransLF}) and (\ref{TransET}),
above] creates a difference in the vertices (and OBE amplitudes)---besides 
the one in the meson propagators.
The $\rho$ includes the tensor coupling
$i\frac{f_\rho}{2M}\sigma_{\mu\nu}q^\nu$ and the LF
OBE amplitude is given in Eq.~(\ref{VOBEv}).
However, in the ET formalism, 
the $\rho$ amplitude consists of Eq.~(\ref{VOBEv}) plus additional
terms that contribute only off-shell, as discussed~\cite{foot1}.
In nuclear matter, the medium effect generated by 
these off-shell terms essentially cancels 
the medium effect
that comes from the main part of the amplitude, Eq.~(\ref{VOBEv}).
Therefore, ET $\rho$ exchange produces a much weaker 
medium effect than LF.

Concerning the pion, 
the pseudovector (pv) coupling
(or gradient coupling) 
$i\frac{f_{\pi NN}}{m_\pi}\gamma_5\gamma_\mu q^\mu$
has been generally used,
in the ET formalism~\cite{rm1}. The resulting
one-pion-exchange (OPE) amplitude can be cast into a form
that consists of the amplitude Eq.~(\ref{VOBEpv}) plus
off-shell terms~\cite{foot2,foot3}. 
In the LF formalism, no matter if the pv or ps coupling is used, 
the OPE amplitude always comes out to be
Eq.~(\ref{VOBEpv}), and there are no additional terms~\cite{foot3}.
In nuclear matter, the medium effect from the off-shell terms 
of the ET formalism damp the medium effect from the main OPE amplitude, 
Eq.~(\ref{VOBEpv}),
similar to what happens with the $\rho$.
Therefore, LF $\pi$ exchange produces a stronger (more repulsive)
medium effect than ET.

In summary, the Dirac effect comes out more repulsive
in the LF formalism as compared to ET due to off-shell differences
in the $\pi$ and $\rho$ exchange amplitudes.
On the other hand, the LF formalism generates an attractive
meson propagator effect that is absent in ET.
As it turns out, these two effects cancel to a large extent, leading to a
non-trivial similarity between 
the LF and ET results.

\section{Full wave function $\Psi$, and meson degrees of freedom}

The nucleonic wave function  $\mid \Phi \rangle $ has been determined in
Eq.~(\ref{Phi}). This gives us the purely nucleonic part of the Fock space,
in which the effects of the mesons have been replaced by the
two-nucleon interaction $K$.
However, the full wave function is 
 $\mid \Psi \rangle $ of Eq.~(\ref{sc}). We need to assess whether
 $\mid \Phi\rangle$
 is a good approximation to $\mid \Psi \rangle $ and we also need
 to determine the mesonic plus-momentum distributions.
 
We recall the  relation between the full $P^-$ operator and the one of
Eq.~(\ref{big}) ($P_0^-=P_0^-(N)+K)$  
corresponding to the nucleonic wave function $\mid \Phi \rangle$:
\begin{equation}
P^- = P^-_0 + J-K+P^-_0(m).
\end{equation}
Using this in Eqs.~(\ref{sc}) and (\ref{big}) allows us 
to  obtain 
\begin{equation}
 \mid \Psi \rangle = \mid \Phi \rangle
 + {1 \over M_A- \Lambda_\Phi P^- \Lambda_\Phi}
 \Lambda_\Phi 
(J-K) \mid \Phi\rangle , \label{psi}
\end{equation}
where $\Lambda_\Phi=1-\mid\Phi\rangle\langle\Phi\mid$.

An expression for the nuclear mass $M_A$ can be obtained by
multiplying Eq.~(\ref{psi}) by
  $\langle \Phi \mid P^-$ 
using 
  $\langle \Phi\mid \Psi\rangle=1$ to obtain the result:
\begin{eqnarray}
M_A = \langle  \Phi \mid P^- \mid \Phi\rangle  + \langle \Phi \mid (J-K)
\Lambda_\Phi 
{1 \over M_A- \Lambda_\Phi  P^- \Lambda_\Phi } \Lambda_\Phi  (J-K) \mid \Phi
\rangle .
\end{eqnarray}
For the purely nucleonic wave function  $\mid \Phi\rangle $ we have
\begin{eqnarray}
\langle  \Phi\mid P^- \mid \Phi\rangle  =
M_0 + \langle \Phi \mid J-K \mid \Phi \rangle,
\end{eqnarray}
so that
\begin{eqnarray}
M_A =M_0+\langle \Phi \mid J-K \mid \Phi\rangle  + \langle \Phi \mid (J-K)
\Lambda_\Phi 
{1 \over M_A- \Lambda_\Phi  P^- \Lambda_\Phi } \Lambda_\Phi  (J-K) \mid \Phi
\rangle .
\end{eqnarray}
The difference between $M_A$ and $M_0$ is the expectation value of the
 operator $O$:
\begin{eqnarray}
O \equiv  J-K + (J-K) \Lambda_\Phi  {1\over M_A-\Lambda_\Phi  P^-_0
  \Lambda_\Phi  - \Lambda_\Phi  
(J-K) \Lambda_\Phi } \Lambda_\Phi  (J-K), 
\end{eqnarray}
and a bit of algebra shows that $O$ satisfies the integral equation
\begin{eqnarray}O = J-K + (J-K) \Lambda_\Phi G_0 (M_A)\Lambda_\Phi O\label{oeq}
\end{eqnarray}
where 
\begin{eqnarray}
G_0(M_A) \equiv  \Lambda_\Phi {1\over M_A-\Lambda_\Phi P^-_0\Lambda_\Phi}
\Lambda_\Phi .
\end{eqnarray}
The lowest relevant order of Eq.~(\ref{oeq}) is given by 
\begin{eqnarray}
O \approx  J-K + (J-K) \Lambda_\Phi G_0 (M_A) (J-K) \nonumber.
\end{eqnarray}

The one-boson exchange interaction   $K$ is given by Eq.~(\ref{kdef0})
(which now includes also the effects of Sect.~III.B) and 
we may determine if the expectation value
$ \langle \Phi \mid O \mid \Phi \rangle$
is reasonably  small. If  this is true,
 the 
quantity $M_0$ would be
 a good approximation to the true eigenvalue of the $P^-$
operator $M_A$.
In the one-boson exchange approximation
\begin{eqnarray}
 \langle \Phi \mid O \mid \Phi  \rangle  &\approx&
 \langle  \Phi \mid v_3-K \mid \Phi \rangle 
+  \langle  \Phi \mid J \,\, G_0 (M_A) J \mid \Phi \rangle \\
&=&  \langle  \Phi \mid  v_3-K \mid \Phi\rangle 
+  \langle  \Phi \mid v_1 G_0 (M_A) v_1 \mid \Phi \rangle. 
\end{eqnarray}
The use of Eq.~(\ref{kdef0}) yields
\begin{eqnarray}
 \langle \Phi \mid O \mid \Phi\rangle &\approx& 
 \langle \Phi \mid v_1\left( G_0 (M_A)-g_0(P^-_{ij})\right) v_1 \mid \Phi
 \rangle ,
\end{eqnarray}
where the term $P^-_{i,j}$ is specified in Eqs.~(\ref{bweinberg})-(\ref{pim}).
But within the independent pair approximation (in which one includes only the
energy (minus-momentum)  differences for a chosen pair of nucleons)
\begin{eqnarray} G_0 (M_A)\equiv g_0(P^-_{ij}),\label{gg}
  \end{eqnarray}
  and 
the expectation value of $O$ vanishes. 

Thus within our approximations, it is consistent to say that
the exact nuclear mass $M_A$ is well
approximated
by $M_0$. This means that we have shown that it is acceptable to remove the
explicit mesons for calculations of the nuclear mass. One could evaluate the
presumably small corrections by going beyond the independent pair
approximation.
The simplicity of the derivation of this result is made possible by the
dynamical simplicity of the vacuum, which is one of the defining features of
light front field theory.

\subsection{Momentum distributions} 
We can now compute the + components of momentum.
Look at $T^{++}$ as given by Eq.~(\ref{tpp}) 
The plus momentum carried by the scalar meson is given by
\begin{eqnarray}
 P^+ (\phi) = \int d^2 k_\perp dk^+ k^+ a^\dagger(k) a(k),
\label{p+phi}\end{eqnarray}
while that of the pion is given by
\begin{eqnarray}
 P^+ (\pi) = \int d^2 k_\perp dk^+ k^+ \bbox{a}^\dagger(k)\cdot\bbox{a}(k),
\label{ppi}\end{eqnarray}
and  that of the vector meson is given by
\begin{eqnarray}
 P^+ (\omega) = \sum_{\omega=1,3} \int d^2 k_\perp dk^+ k^+ a^\dagger 
({\bbox k}, \omega) a ({\bbox k}, \omega) . \end{eqnarray}

We shall handle the 
scalar term first. The evaluation of (\ref{p+phi}) using 
(\ref{psi}) leads to 
\begin{eqnarray}
 P^+(\phi)& =& \int d^2 k_\perp dk^+ k^+ \langle\Phi \mid j ( {\bbox k} 
) {1\over (M_A-\Lambda P^- \Lambda)^2} j ({\bbox k}) \mid \Phi\rangle\nonumber\\
&\approx &\int d^2 k_\perp dk^+ k^+ \langle\Phi \mid j ( {\bbox k} 
) {1\over (M_A-\Lambda P^-_0 \Lambda)^2} j ({\bbox k}) \mid \Phi\rangle,
\label{ppp}
\end{eqnarray}
in which the approximation is motivated by the near equality of
$M_A$ and $M_0$ and the universal expectation that the impulse approximation
evaluation of the meson exchange potential is valid.
The term $j(k)$ is defined via the contribution of the
scalar mesons to $v_1$:
\begin{eqnarray}
 v_1(\phi) = \int d^2 k_\perp dk^+ [j ({\bbox k}) a ({\bbox k}) +H.C],
\end{eqnarray}
in which $j ({\bbox k})$ can be obtained from Eq.~(\ref{v1}) and is a nucleonic
operator that depends on $\bbox{k}_\perp$ and $k^+$.  

The operator to be
evaluated in the above equation has one and two body pieces. The one-body
terms are related to a shift in the self energy of the nucleon caused by
the medium. In
infinite nuclear matter the ratio of pairs to single nucleons is infinite
so that the number density is well approximated by the two nucleon terms of
Eq.~(\ref{ppp}). The evaluation is simplified by the use of Eq.~(\ref{gg}) and
noting that the relevant matrix element is the same as occuring in the
one-boson exchange operator $K$ except that the denominator is squared.
Thus the momentum density $n_\phi({\bbox k})$, defined by
\begin{eqnarray}
 P^+ (\phi) \equiv \int d^2 k_\perp dk^+ k^+n_\phi({\bbox k}),
\end{eqnarray}
is given as a derivative of the scalar-meson exchange
contribution to the nucleon-nucleon potential:
\begin{eqnarray}
n_\phi({\bbox k}) 
\approx 
 -2\langle\Phi \mid
 {\partial V_\phi(P^-_{ij}(\bbox{k})) \over \partial P^-_{ij} }
 \mid \Phi \rangle 
\end{eqnarray}
with
\begin{eqnarray}
  {\partial V_\phi(P^-_{ij}(\bbox{k})) \over \partial P^-_{ij} }
\equiv \left[ j_i ( {\bbox k} 
) {1\over (M_A-\Lambda P^- \Lambda)^2} j_j ({\bbox k})\right] 
\end{eqnarray}
  in which the notation $i,j$ specifies that only two-nucleon contributions are
  included. Note that 
$\mid \Phi >$ is the correlated ground state. Note that the expectation value
is taken using the single particle basis specified by Eq.~(\ref{alphas}).
We recall Eqs.~(\ref{kdef0}) and (\ref{gij}), and use\cite{gam97b}
\begin{equation}
k^+(P_{ij}^--P_0^-) =q^2-m_\phi^2.
\end{equation}
Note that the momentum of the exchanged meson is k, and
\begin{eqnarray}k^+=q^+,\bbox{k}_\perp=\bbox{q}_\perp,
  k^-={\bbox{k}_\perp^2+m_\phi^2\over k^+},
  \end{eqnarray}
  where $q$ is the nucleon momentum transfer. Thus one may obtain
the result that 
\begin{equation}
  -{\partial \over \partial P_{ij}^-} {1\over (P_{ij}^--P_0^-)}=
   { k^+\over(q^2-m_\phi^2)^2}.
    \end{equation}
    This means that evaluating the  plus-momentum  distribution for scalar
    mesons is the same as evaluating the expression for the scalar meson
    contribution to the nuclear potential energy, except that the
    potential is multiplied by the factor $ -k^+/(q^2-m_\phi^2)$.
    The net result is that
    \begin{eqnarray}
     n_\phi({\bbox k})=\sum_{\alpha,\beta<F} \langle \alpha\beta \mid
     \Omega^\dagger_{\alpha\beta}V_\phi({\bbox k}){(- k^+)\over(q^2-m_\phi^2)}
    \Omega_{\alpha\beta} \mid \alpha\beta\rangle_a,
     \end{eqnarray}
     where $\Omega_{\alpha,\beta}$ is the Moeller scattering operator for the
     two-nucleon state $ \alpha\beta$. One finds a similiar expression for
     the pionic density with
  \begin{eqnarray}
     n_\pi({\bbox k})=\sum_{\alpha,\beta<F} \langle \alpha\beta\mid
     \Omega^\dagger_{\alpha\beta} V_\pi({\bbox k}){(- k^+)\over(q^2-m_\pi^2)}
    \Omega_{\alpha\beta} \mid \alpha\beta\rangle_a.
     \end{eqnarray}

     The evaluation of the vector meson  density $n_\omega({\bbox k})$
     requires more  steps because the meson-exchange potential has a
     contribution from the instantaneous meson exchange term $v_3$. This
      instantaneous term does not lead to a meson ``in the air'' and therefore
      does not contribute  to $n_\omega({\bbox k})$. One finds that
  \begin{eqnarray}
     n_\omega({\bbox k})=\sum_{\alpha,\beta<F} \langle \alpha\beta\mid
     \Omega^\dagger_{\alpha\beta}\tilde{ V}_\omega
     ({\bbox k}){(- k^+)\over(q^2-m_\omega^2)}
    \Omega_{\alpha\beta} \mid \alpha\beta\rangle_a,
     \end{eqnarray}
     where
     \begin{eqnarray}
      \tilde{ V}_\omega({\bbox k})
       = g_\omega^2/(2\pi)^3
\frac{F^2_\omega(q^2)} 
{\left(q^2-m_v^2\right)} 
\left[{ k_\perp^2+m_v^2\over {k^+}^2}\right]. 
\label{vvex}
\end{eqnarray}

It is also necessary to discuss the nucleonic plus-momentum distribution. This
is determined in Ref.~\cite{gam97b}. Here the nucleon-nucleon correlations
cause the momentum density to have contributions from above the 
    Fermi sea.
    We have 
\begin{equation}
{P^+_N\over A}={4\over\rho_B (2\pi)^3}\int d^2k_\perp dk^+ k^+ 
N(k_\perp,k^+),\label{nuc}
\end{equation} 
where $\rho_B$ is the nuclear baryon density and $N(k_\perp,k^+)$ is the
occupation number for a nucleon of momentum $(k_\perp,k^+)$.
Recall that
the variable $k^+$ is defined in Eq.~(\ref{kplus}).
Since the integral gives the total plus-momentum
carried by nucleons, the integrand (over $k^+$) which multiplies the
factor $k^+$ is  the  probability $f(k^+)$ that a nucleon has momentum $k^+$
distribution. Thus:
\begin{equation}{P^+_N\over A}=\int
dk^+\;k^+ f(k^+),
\end{equation}
where
\begin{equation}
  f(k^+)= {4\over\rho_B (2\pi)^3}\int d^2k_\perp N(k_\perp,k^+).
  \end{equation} 
A  function $f(y)$ can be obtained by 
replacing $k^+$ by the dimensionless variable $y$ using
$y\equiv {k^+\over \bar{M}}$
with $\bar{M}\equiv M-14.71 $ MeV.

The  relation to experiments is obtained by
 recalling that the nuclear structure function $F_{2A}$ 
can be obtained from the light front distribution function
$f(y)$ (which gives the probability for a nucleon to have 
a plus momentum fraction $y$) and the nucleon structure function
$F_{2N}$ using  the relation\cite{sfrel}:
\begin{equation}
{F_{2A}(x)\over A}=\int dy f(y) F_{2N}(x/y), \label{deep}
\end{equation}
where $y$ is $A$ times the fraction of the 
nuclear  plus-momentum carried by the nucleon, and
$x$ is the Bjorken variable computed using the
nuclear mass minus the binding energy.
This formula  is the expression of the usual convolution model, with
validity determined by a number of assumptions.
Our formalism enables us to calculate
the function $f(y)$ from the integrand of 
Eq.(\ref{pplus}).

\subsection {Computing the total number of mesons} 

We can get the total number of each kind of meson
(except the $\omega)$ using a sum rule.
Consider the schematic form of the equation for the G-matrix
\begin{equation}
  G(P_{ij}^-)=V(P_{ij}^-) +V(P_{ij}^-){Q\over \Delta E} G(P_{ij}^-),
  \label{gmats}
  \end{equation}
 where
 ${Q\over \Delta E} $ is a schematic
 representation of the propagator of Eq.~(\ref{bweinberg}).
   Differentiating with respect to
 $P_{ij}^-$ yields the result:
\begin{equation}
  {\partial G(P_{ij}^-)\over \partial P_{ij}^-}=
  (1+G{Q\over \Delta E})  {\partial V(P_{ij}^-)\over \partial P_{ij}^-}
(1 +{Q\over \Delta E}G). 
\end{equation}
The Moeller operators appear to the right and left of the derivative of the
potential. Furthermore, in our one boson exchange approximation, the total
nucleon-nucleon potential is the sum of the contributions due to individual
bosons. Thus we may define
\begin{equation}
  {\partial G^m(P_{ij}^-)\over \partial P_{ij}^-}\equiv
  (1+G{Q\over \Delta E})  {\partial V^m(P_{ij}^-)\over \partial P_{ij}^-}
(1 +{Q\over \Delta E}G), 
\end{equation}
in which the label $m$ refers to the type of meson. But
The potential $V^m$ appearing in Eq.~(\ref{gmats}) is simply
the Fourier transform
of the potential $V^m(\bbox{q})$. Thus  
an examination of eqns. (6.25) and (6.24) shows that, considering the pion
for example, 
\begin{equation}
  N_\pi\equiv \int d^3q n_\phi(\bbox{q})=\sum_{\alpha,\beta<F}
  \langle \alpha,\beta\mid  {\partial G^\pi(P_{ij}^-)\over \partial P_{ij}^-}
  \mid \alpha,\beta\rangle_A. \label{npi}
  \end{equation}
  Numerical evaluation of Eq.~(\ref{npi})
  leads to the result that $N_\pi/A= 0.05.$
  This is smaller than the 18\% of Friman et al\cite{bf}
because we use scalar mesons  instead of intermediate $\Delta$ states to 
 provide the bulk of the attractive 
force.

The expression for the density of vector mesons involves the removal of
  the effects of the instantaneous exchange
  and one must use explicit light front
  variables.

\section{Implications for  Lepton-Nucleus Deep Inelastic Scattering and
  the nuclear Drell-Yan Process}

The values of $U_S$, $U_V^-$ and $N_\pi$ allow us to assess the 
deep inelastic scattering of leptons from our version of the ground state of
nuclear matter. Using $M^*(k_F)=744$ MeV~\cite{foot4} and 
neglecting the influence of
two-particle-two-hole states
to approximate
$f(k^+)$\cite{gam97a,gam97b,slater} 
shows that nucleons carry 81\% (as opposed to the 65\% 
of mean field theory\cite{gam97a}) of  the nuclear plus momentum. 
This represents a vast improvement in the description of nuclear deep inelastic
scattering. The 
minimum value of the ratio $F_{2A}/F_{2N}$, obtained from 
the convolution formula (\ref{deep}) 
 is increased by a factor of twenty   towards 
the data 
as  
extrapolated in Ref.~\cite{Sick}. But this calculation provides only  a
lower limit of the nucleon contribution because of the neglect of 
effects of the two-particle-two hole states\cite{foot5}.

Turn now to the experimental information about the nuclear pionic content.
The Drell-Yan experiment on nuclear targets\cite{dyexp}
showed no enhancement of nuclear pions within an error of about 5\%-10\% for
their heaviest target. No substantial  pionic enhancement 
is found in (p,n) reactions\cite{pn}.
Understanding this result is 
an important challenge to the 
understanding of nuclear dynamics~\cite{missing}. 
Here we have a good description of nuclear dynamics, 
and our 5\%  enhancement is consistent\cite{jm},
within errors, with the Drell-Yan
data.

\section{Summary and Discussion}

This paper contains a new 
relativistic light-front theory of nuclear matter. Light front quantization is
used to obtain a  nucleon-nucleon potential which yields phase shifts in good
agreement with data.  We use this as input in a light front many body theory.
A straightforward derivation leads to a theory
in which the effective interaction is
the light-front G-matrix. We obtain a good
description of the binding energy, density, and 
incompressibility of nuclear matter. The binding energy per nucleon
is 14.7 MeV and $k_F=1.37 $fm$^{-1}$. The compressibility is 180 MeV.

The use of a meson-nucleon Lagrangian enables us to also compute the 
mesonic content of the  wave function using a consistent approach represented
by Eqs.~(\ref{sc}),(\ref{big}) and (\ref{psi}).
  The results are not in conflict
with extrapolations of deep inelastic scattering and Drell-Yan data to 
nuclear matter.

The omega and rho mesonic content are still to be evaluated. We believe
that the possible  
nuclear enhancement of vector mesons is a promising avenue for future
theoretical and experimental research.

\section*{Acknowledgments}
This work was supported in part the the U.S.\ National Science
Foundation under Grant No.\ PHY-9603097 and by the U.S.D.O.E.

%
%
%
%
\appendix
\section{Notation, conventions, and useful relations}
\noindent This is patterned after the review of Harindranath\cite{hari}
 The  light-front variables are defined by
\begin{equation} x^{+}= x^{0}+x^{3} \; , \; \;\; x^{-}= x^{0}-x^{3},
\label{def}
 \end{equation}
so  the four-vector $x^\mu$ is denoted
\begin{equation} x^{\mu} = (x^{+},x^{-},\bbox{x}^{\perp}) . \end{equation}
With this notation the 
scalar product is denoted by 
 \begin{equation} x\cdot y = {1 \over 2} x^{+}y^{-}+{1 \over 2}x^{-}y^{+}-
\bbox{x}^{\perp}\cdot \bbox{y}^{\perp}  . \end{equation}
The metric tensor $g^{\mu \nu}$ with $\mu=(+,-,1,2)$ is obtained from the 
usual one by using (\ref{def}) (i.e. $g^{0\mu}=g^{0\mu}+g^{3\mu}$). Then 
$g^{+-}=g^{-+}=2,g^{ij}=-1$, with the other elements vanishing.
The term $g_{\mu\nu}$ is obtained from the condition  that 
$g^{\alpha\beta}g_{\beta\gamma}=\delta_{\alpha\gamma}$. Its elements are the
same as those of $g^{\mu\nu}$ except for $g_{-+}=g_{+-}=1/2$.
Thus
  \begin{eqnarray}
 x_{-}= {1 \over 2} x^{+}  , \; \; x_{+} = {1 \over 2} x^{-}, \end{eqnarray}
\noindent and the  partial derivatives are similarly given by 
\begin{eqnarray}
\partial^{+}= 2 \partial_{-}= 2 {\partial \over \partial x^{-}} \;\quad
\partial^{-}= 2 \partial_{+}= 2 {\partial \over \partial x^{+}}  . 
\end{eqnarray}

The Bjorken and Drell\cite{Bj}  convention for gamma matrices is used
and 
\begin{eqnarray} 
\gamma^{\pm} \; \; \equiv \gamma^{0} \pm \gamma^{3}  . \end{eqnarray}
the relations 
\begin{equation}\gamma^\pm\gamma^\pm=0,
\quad\gamma^+\gamma^-\gamma^+=4 \gamma^+,\quad 
\gamma^-\gamma^+\gamma^-=4 \gamma^-
\end{equation} can be used to simplify various computations.

The Hermitian projection operators $\Lambda_\pm$ are given by 
\begin{eqnarray}
  \Lambda_{\pm}  \; =  \; {1 \over 4} \gamma^{\mp} \gamma^{\pm}
 = {1 \over 2} \gamma^{0} \gamma^{\pm} \; = \; {1 \over 2}
(I \pm \alpha^{3})  , \end{eqnarray}
and obey the following relations
\begin{eqnarray} (\Lambda_{\pm})^2  \; \; = \; \; \Lambda_{\pm},\quad \quad
\gamma^{\perp} \; \Lambda_{\pm}, \; \; 
= \; \; \Lambda_{\pm} \gamma^{\perp}  , \end{eqnarray}
\begin{eqnarray}
  \gamma^{0} \; \Lambda{\pm} \; \; = \; \; \Lambda_{\mp} \gamma^{0}
\quad\quad
           \alpha^{\perp} \; \Lambda_{\pm} \; \; 
= \; \; \Lambda_{\mp} \alpha^{\perp} , \end{eqnarray}
\begin{eqnarray}  \gamma^{5} \; \Lambda_{\pm} \; \; 
= \; \; \Lambda_{\pm} \gamma^{5} \quad
 \gamma^\mp = 2 \Lambda_\pm \gamma^0 = \gamma^\mp \Lambda_\mp ,\end{eqnarray}
\begin{eqnarray}
 \gamma^i \Lambda_\mp = {1 \over 2} \gamma^i \pm i {1 \over 2} \epsilon^{ij}
\gamma^j \gamma^5 , \end{eqnarray}
\begin{eqnarray}
 \alpha^j \gamma^i \Lambda_+ = {i \over 2} \epsilon^{ij} \gamma^+ \gamma^5
. \end{eqnarray}   

\pagebreak

\begin{table}
\caption{Potential parameters and predictions for the deuteron
and low-energy $np$ scattering.
For the deuteron, the binding energy $B_d$, the $D$-state probability $P_D$,
the quadrupole moment $Q_d$, and the asymptotic $D$-state over $S$-state
ratio $D/S$ are given. Low-energy $np$ scattering is parametrized in terms
of $a_{np}$ and $r_{np}$ in $^1S_0$ and $a_t$ and $r_t$ in $^3S_1$,
where $a$ denotes the scattering length and $r$ the effective range.
The nucleon mass is $M=938.919$ MeV.}
\begin{tabular}{lllllllll}
  &            & \multicolumn{2}{c}{\bf Light-Front OBEP}
&\hspace*{.1cm}& \multicolumn{2}{c}{{\bf Thompson OBEP}$^a$}
&\hspace*{.1cm}& \multicolumn{1}{c}{{\bf Empirical}$^b$} 
\\ \cline{3-4} \cline{6-7} \cline{9-9} \\
\multicolumn{9}{l}{\rm Meson Parameters:}\\
\\
  & $m_{\alpha}$(MeV)
     &  $g^{2}_{\alpha}/4\pi$ $[f/g]$
     &  $\Lambda_{\alpha}$(GeV)
     &&  $g^{2}_{\alpha}/4\pi$ $[f/g]$
     &  $\Lambda_{\alpha}$(GeV)
     &&  $g^{2}_{\alpha}/4\pi$ $[f/g]$ 
\\ \hline 

$\pi$ & 138.04  & 14.0 & 1.2 && 14.6 & 1.2 && 13.5 -- 14.6 \\

$\eta$ & 547.5  &  3   & 1.5 &&   5  & 1.5 && $\leq 5$ \\

$\rho$ & 769 & 0.9  [6.1] &1.85&& 0.95 [6.1]& 1.3 && 0.6(1) $[6.6\pm 1.0]$\\

$\omega$ & 782 & 24.5 [0.0] & 1.85 && 20.0 [0.0] & 1.5 && $24\pm 5\pm 7$ \\

$a_0$ & 983  & 2.0723 & 2.0 && 3.1155 & 1.5 &&              \\

$\sigma$ & 550  & 8.9602 & 2.0 &&  8.0769  &  2.0 &&  
\\ \hline \\
\multicolumn{9}{c}{\rm Deuteron}\\
\\
 $B_{d}$ (MeV)&&  2.2245 &&& 2.2247 &&& 2.224575(9)\\
 $P_{D}$ (\%) && 4.53 &&& 5.10 &&&  ---\\
 $Q_{d}$ (fm$^{2}$) && 0.270$^{c}$ &&& 0.278$^{c}$ &&&  0.2860(15)\\
 $D/S$ &&   0.0250 &&&   0.0257  &&&  0.0256(4)\\ 
\\
\multicolumn{9}{c}{\rm Low-Energy $np$ Scattering}\\
\\
 $a_{np}$ (fm) && --23.745  &&& --23.747 &&& --23.748(10)\\
 $r_{np}$ (fm) && 2.671    &&&  2.664   &&& 2.75(5)\\
 $a_{t}$ (fm)  && 5.494    &&&  5.475   &&& 5.424(4)\\
 $r_{t}$ (fm)  && 1.856    &&&  1.828   &&& 1.759(5)
\end{tabular}
$^{a}$Potential B of Brockmann and Machleidt~\cite{rm1}.\\
$^{b}$For more comprehensive information on the empirical data and references,
see Table 4.1 and 4.2 of Ref.~\cite{Mac89}.\\
$^{c}$Meson-exchange current contributions not included.
\end{table}

\pagebreak

\begin{center}
FIGURES
\end{center}

\begin{figure}
{FIG.~1. Phase shifts $\delta$ and mixing parameters $\epsilon$
of neutron-proton scattering for partial waves with $J\leq 2$
and laboratory kinetic energies $T_{lab}\leq 300$ MeV.
The solid line is the prediction by the LF OBEP presented in Sec.\
III and the dotted line the one by Potential B of 
Brockmann and Machleidt~\cite{rm1}. The open circles represent
the multi-energy $np$ analysis by the Nijmegen group~\cite{Sto93}
and the solid dots are the VPI analysis SM97~\cite{VPI97}.}
\end{figure}

\begin{figure}
{FIG.~2. Energy per nucleon in nuclear matter, E/A (in units of MeV),
as a function of density expressed by the Fermi momentum $k_F$
(in units of fm$^{-1}$). The solid line is our prediction using
light-front Brueckner theory. The dotted curve is obtained
when the medium effect on meson propagation is omitted.
The dashed line is the result from conventional (nonrelativistic)
Brueckner theory. The box describes the area in which 
nuclear saturation is expected to occur empirically.}
\end{figure}


\begin{references}
\bibitem{notation} Our notation is that a four vector $A^\mu$ is defined by
the plus, minus and perpendicular components as
$(A^0+A^3,A^0-A^3,\bbox{A}_\perp)$ and $A\cdot B={1\over 2} A^+B^-+{1\over
2}A^-B^+-\bbox{A_\perp\cdot B_\perp}$. See the Appendix for more on this.
\bibitem{scalar} The use of scalar mesons is meant as a simple  representation
of part of the two pion
exchange potential which causes much of the  
the medium range attraction between nucleons, as well as the effects of 
a possible fundamental scalar meson. 
\bibitem{difference} If one re-writes the Weinberg-type equation in terms of
  equal time variables, the result is the same as the Blancebecler-Sugar
  equation except for the appearance in the meson propagator
  of the energy difference between on-shell
  initial and final nucleons.
\bibitem{foot1}
Note that the explicit expression for the tensor coupling OBE amplitude
in the LF formalism differs from the one used in the BbS formalism
[compare Eq.~(A.9) of Ref.~\cite{Mac89} or Eq.~(1.61) of
Ref.~\cite{Mac93} with our Eq.~(\ref{VOBEv})]. The reason for this difference
is that in the BbS formalism $q_0=0$ while in the LF formalism
$q_0=E'-E$. As a consequence of this,
the partial-wave decomposed tensor expressions
as given in Ref.~\cite{Mac93} do not apply here.
The expressions to be used here can be found in Ref.~\cite{HM75}
(Appendix A, case 1, therein).
Moreover, all meson propagators 
$[m^2_\alpha + (\bbox{k}'-\bbox{k})^2]^{-1}$
in Ref.~\cite{Mac93} must be replaced by
$[m^2_\alpha -(E'-E)^2 + (\bbox{k}'-\bbox{k})^2]^{-1}$;
and a similar replacement must be done in the denominators
of the cutoffs.
\bibitem{foot2}
The explicit expression for the pv OPE amplitude as derived in
the ET formalism is given
in Eq.~(A.7) of Ref.~\cite{Mac89} and Eq.~(1.59) of
Ref.~\cite{Mac93}.
\bibitem{foot3}
When comparing OPE amplitudes using different couplings,
the coupling constants must be related properly.
The relationship is
\begin{displaymath}
g_\pi = \left( \frac{2M}{m_\pi} \right) f_{\pi NN}
\end{displaymath}
in free space and
\begin{displaymath}
g_\pi^* = \left( \frac{2M^*}{m_\pi} \right) f_{\pi NN}
\end{displaymath}
in the nuclear medium.
The latter is related to the fact
that, based upon our Lagrangian Eq.~(\ref{lag}),
the ps pion coupling constant in the medium is
\begin{displaymath}
g_\pi^* = \frac{M^*}{f} = \frac{M^*}{M} \frac{M}{f}
= \frac{M^*}{M} g_\pi
\end{displaymath}
with $g_\pi=M/f$ the free-space coupling constant
($f$ is the pion decay constant).
\bibitem{sfrel} Eq.~(\ref{deep}) is essentially Eq. (5.2) of 
Frankfurt and Strikman, \cite{fs2}, with a correspondence
between our $A\;f(y)$ 
and their $\int d^2p_\perp \rho^N_A(\alpha,p_\perp)/\alpha$.
\bibitem{foot4} $U$ depends on momentum, so to use 
 a constant value is to take an average.
 Eqs.~(\ref{bbsbs}) and (\ref{U1})
are non-linear, and the solutions have 
a  spread of values. Indeed, 
 $M^*(k_F)= 744$ also  leads to good saturation properties.
\bibitem{slater} We use the Slater determinant $\mid \phi\rangle$ to evaluate
  the expectation value of the nucleon number operator. This means that the
 occupation number $N$ of  Eq.(\ref{nuc}) is either unity or zero.
\bibitem{foot5} We estimate  that
nucleons with momentum greater than $k_F$ would substantially increase the
computed ratio $F_{2A}/F_{2N}$ because $F_{2N}(x)$ decreases very rapidly
with increasing values of $x$ and because $M^*$ would increase at high momenta.
\end{references}
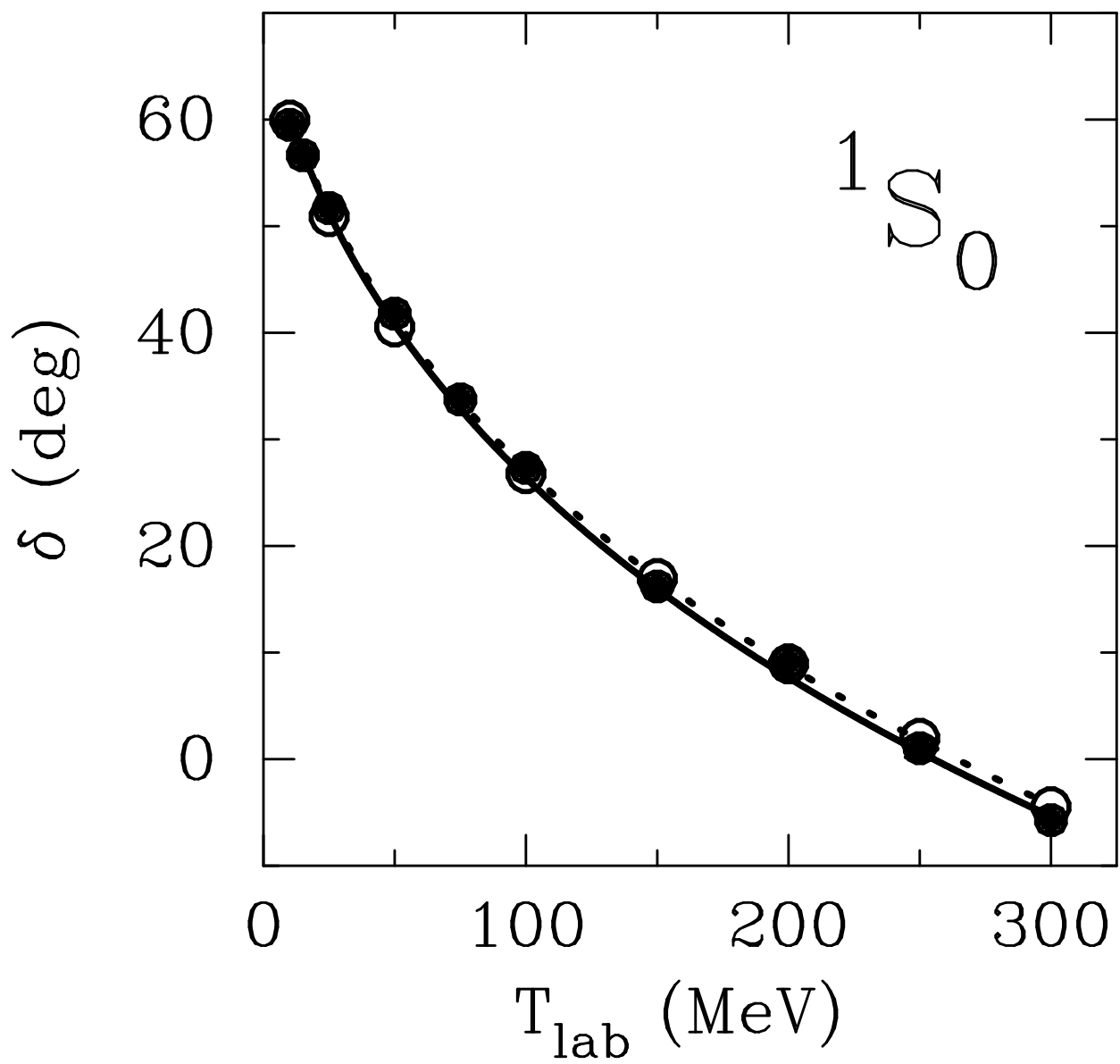
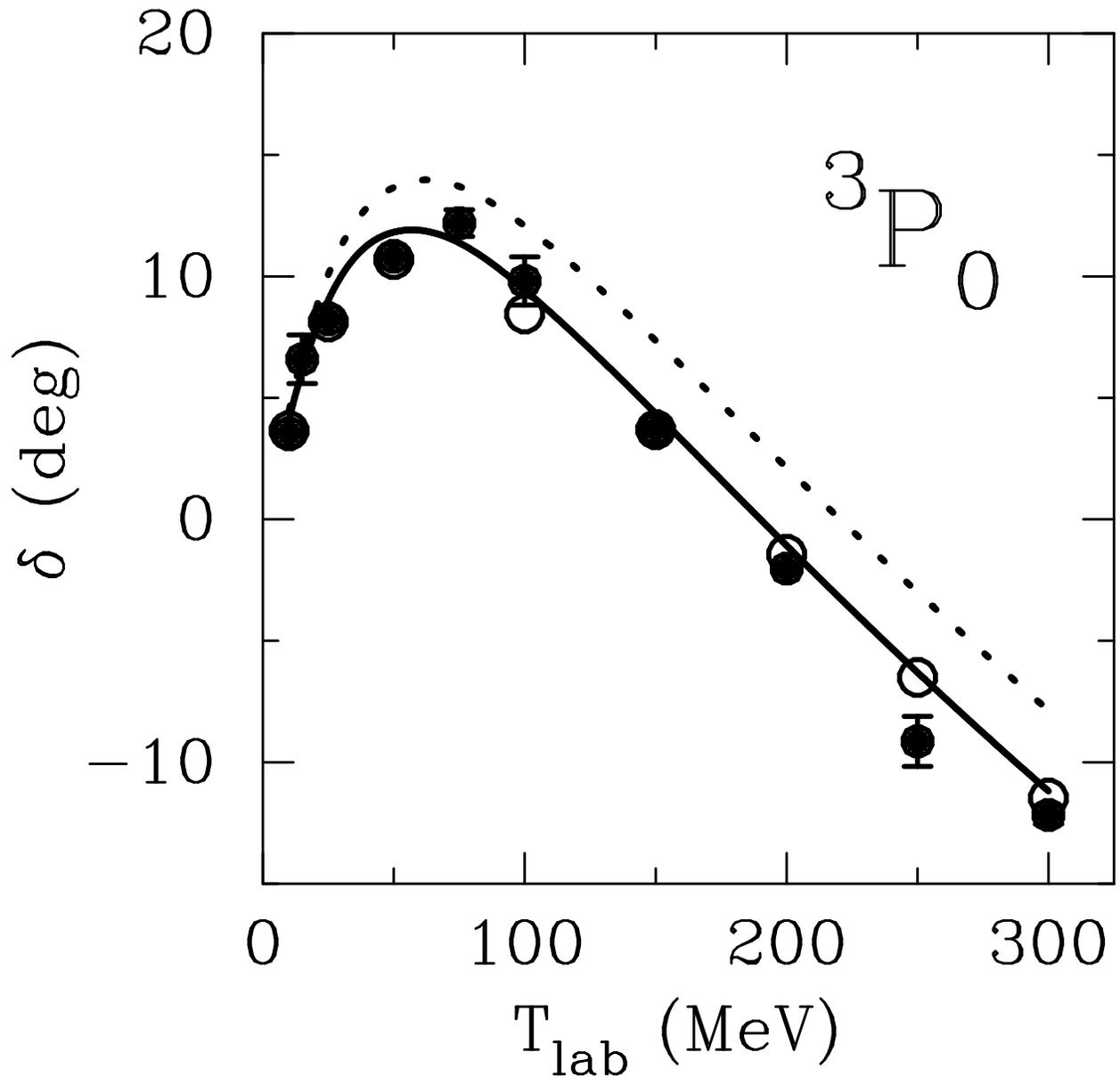
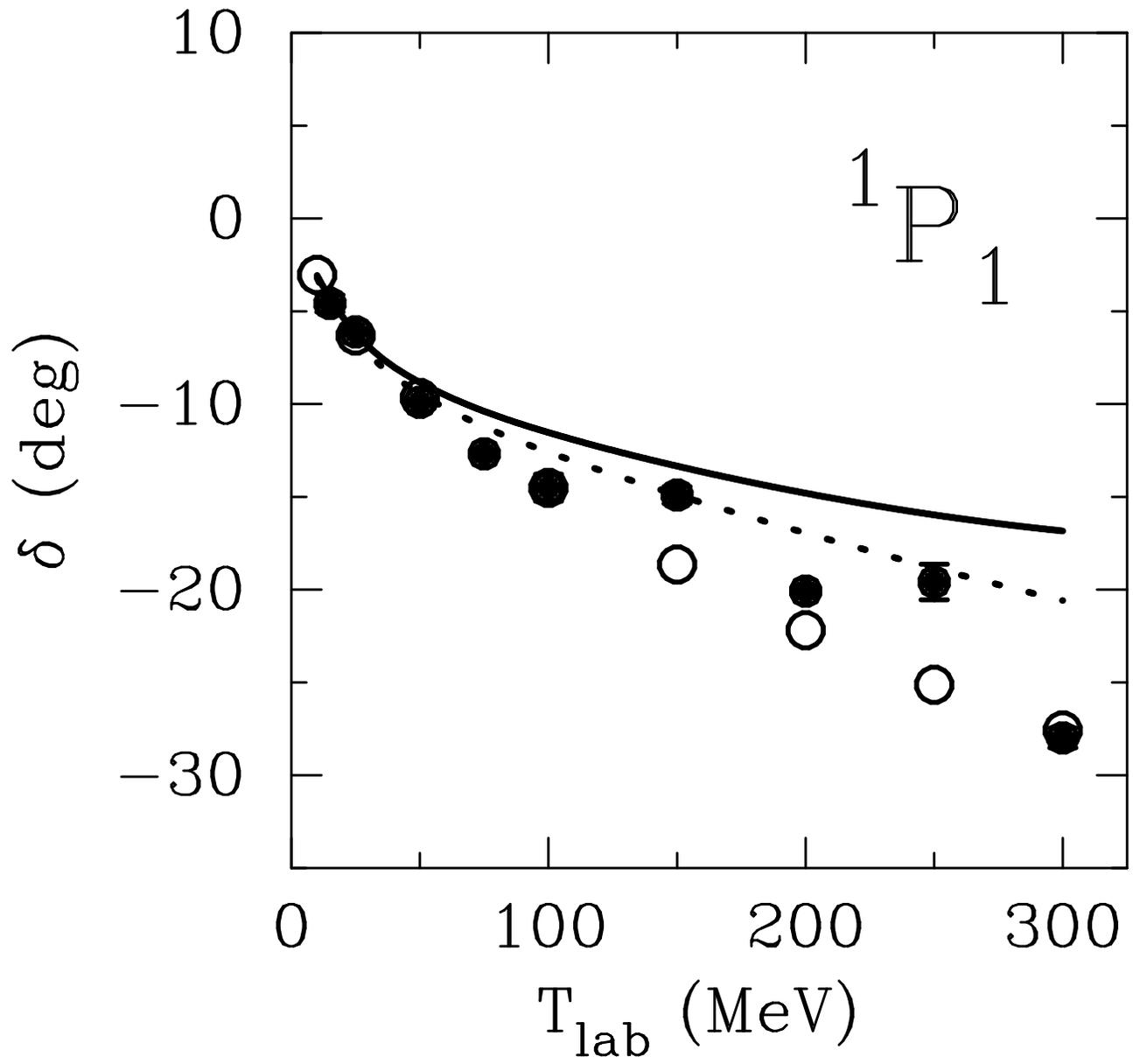
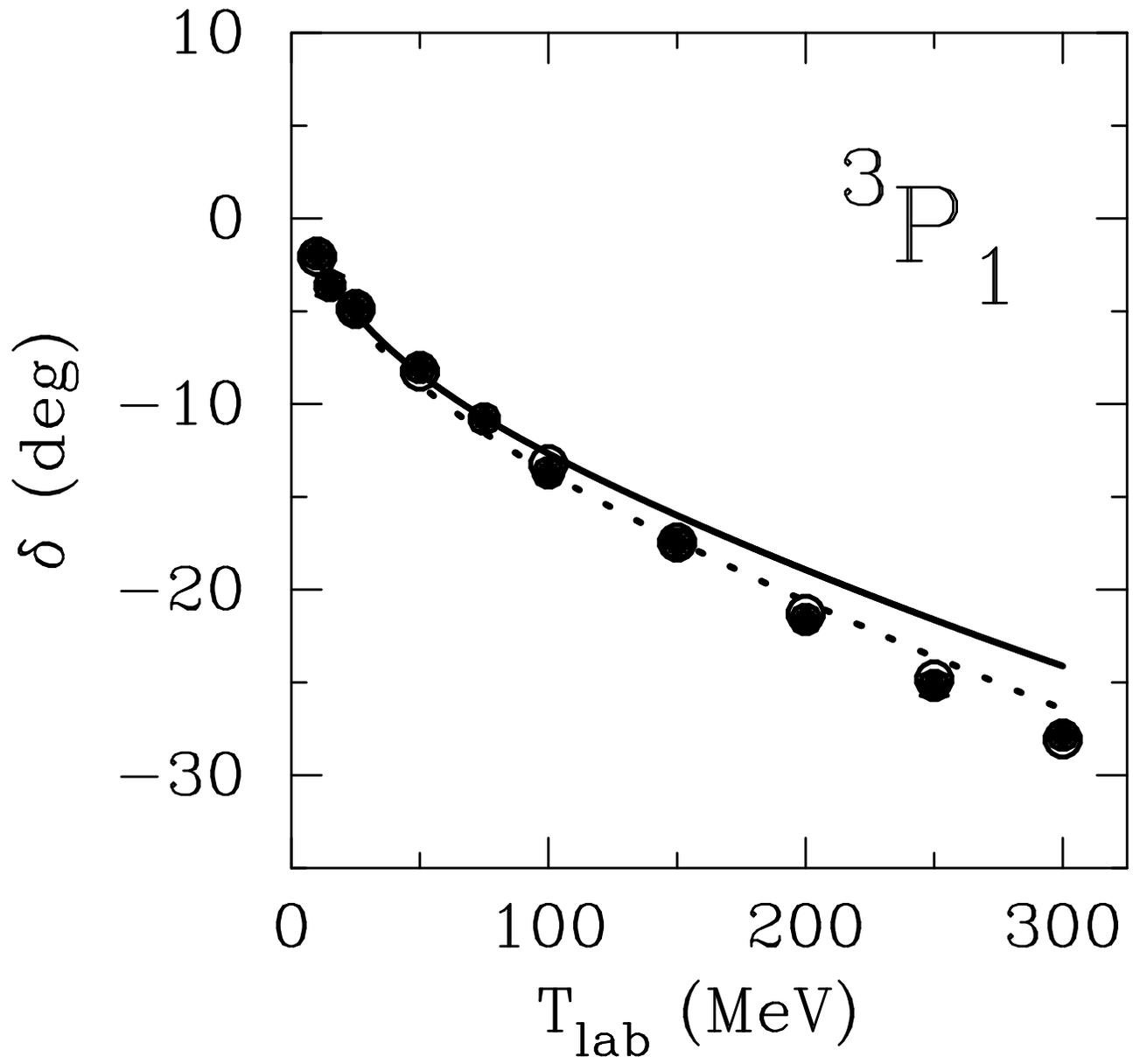
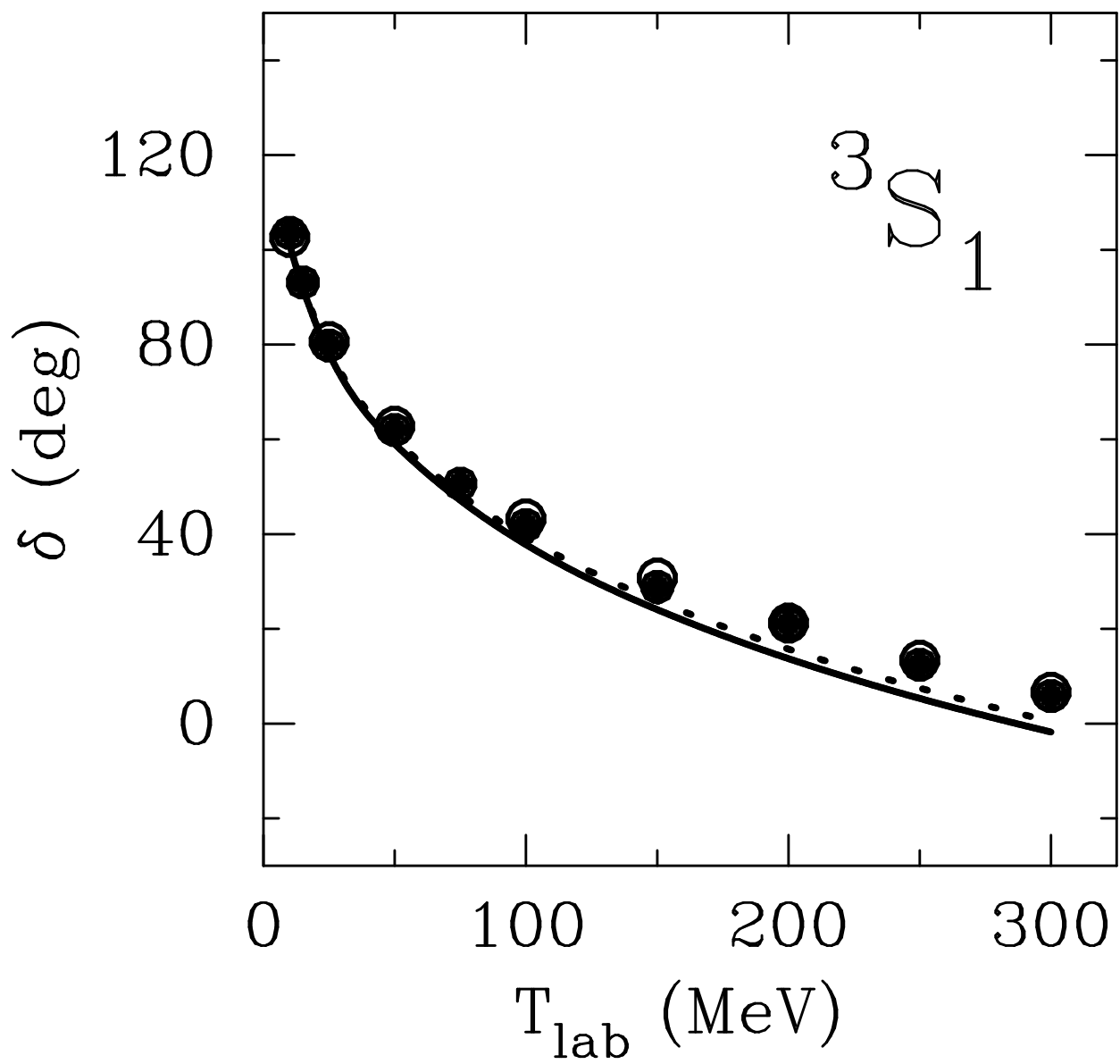
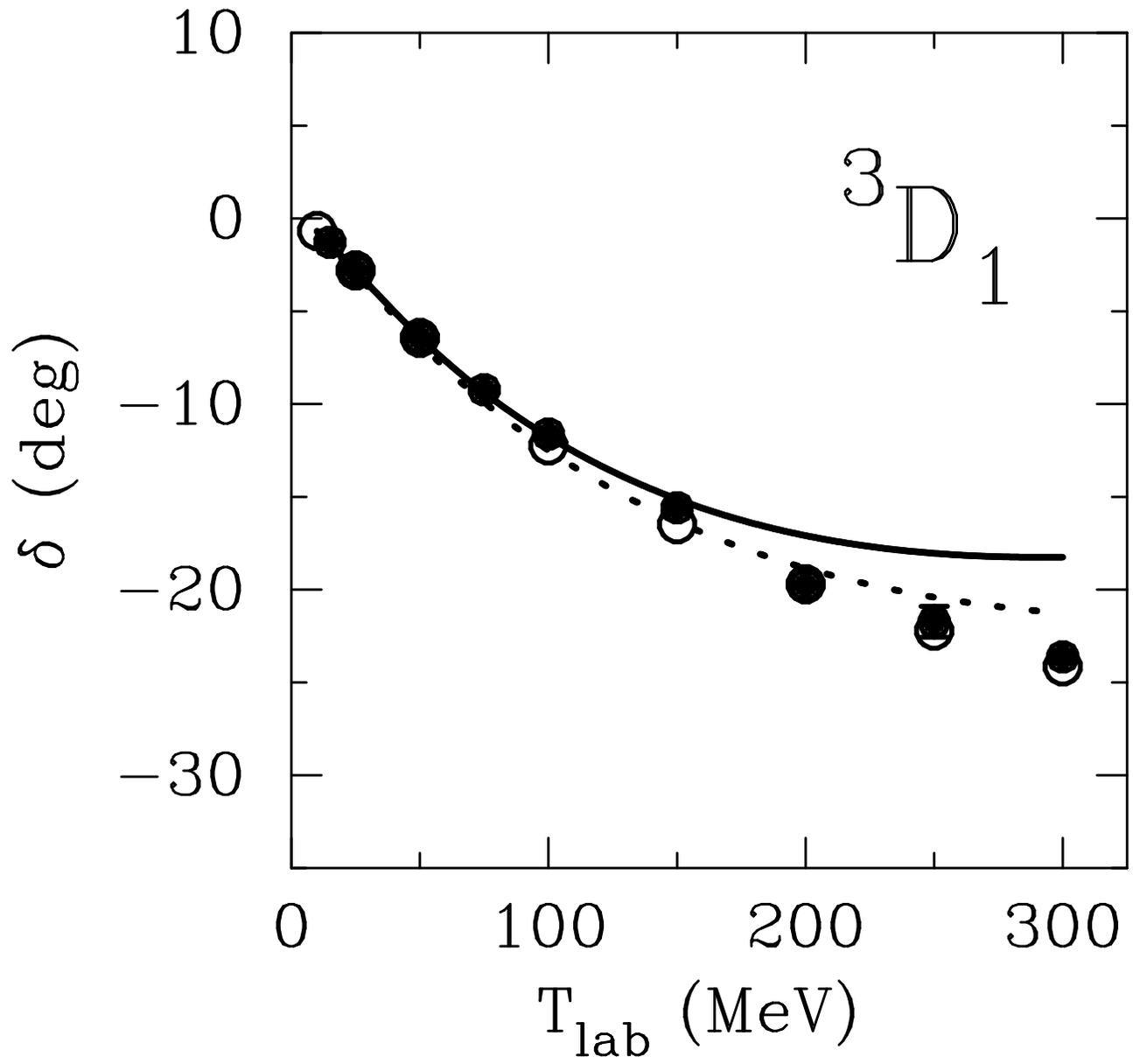
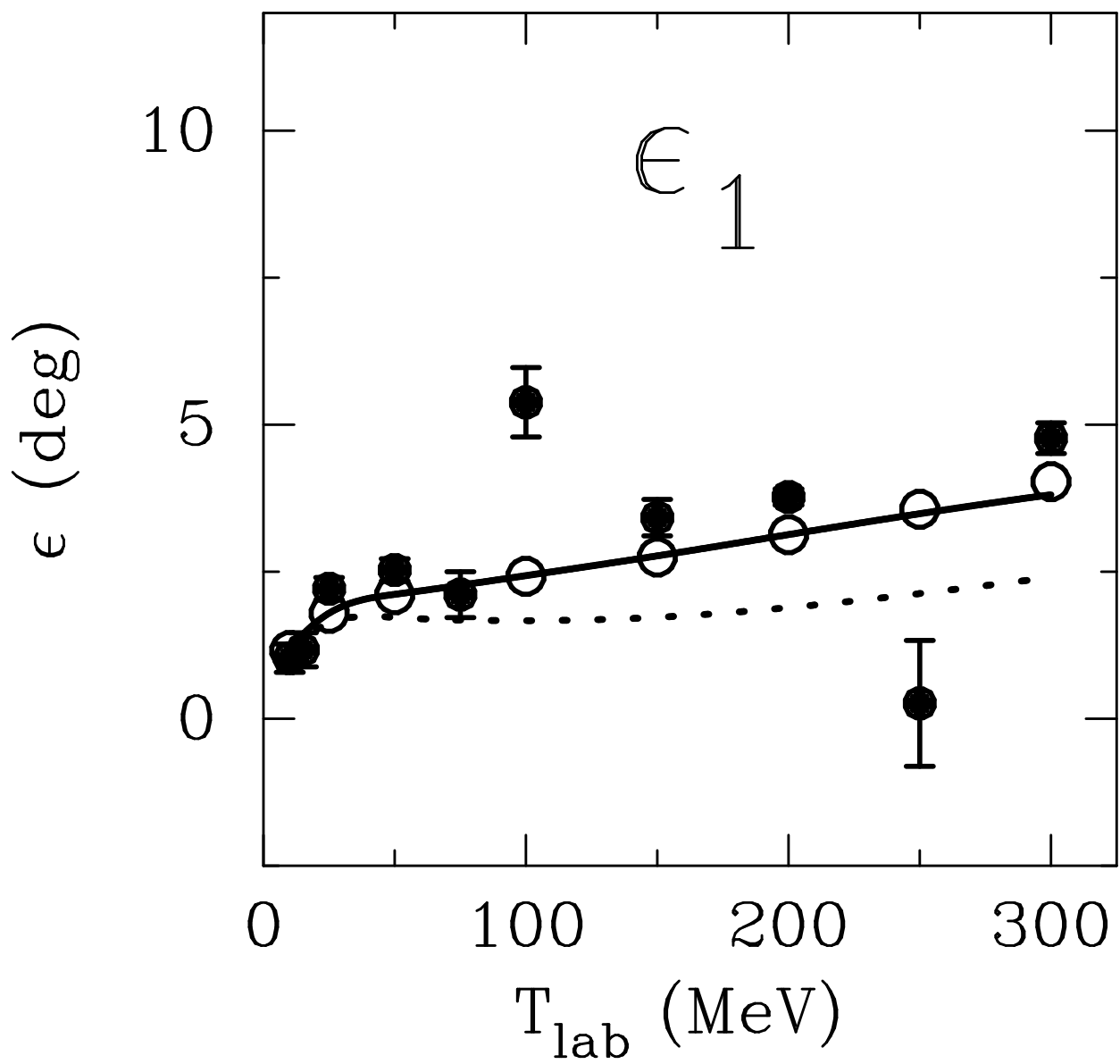
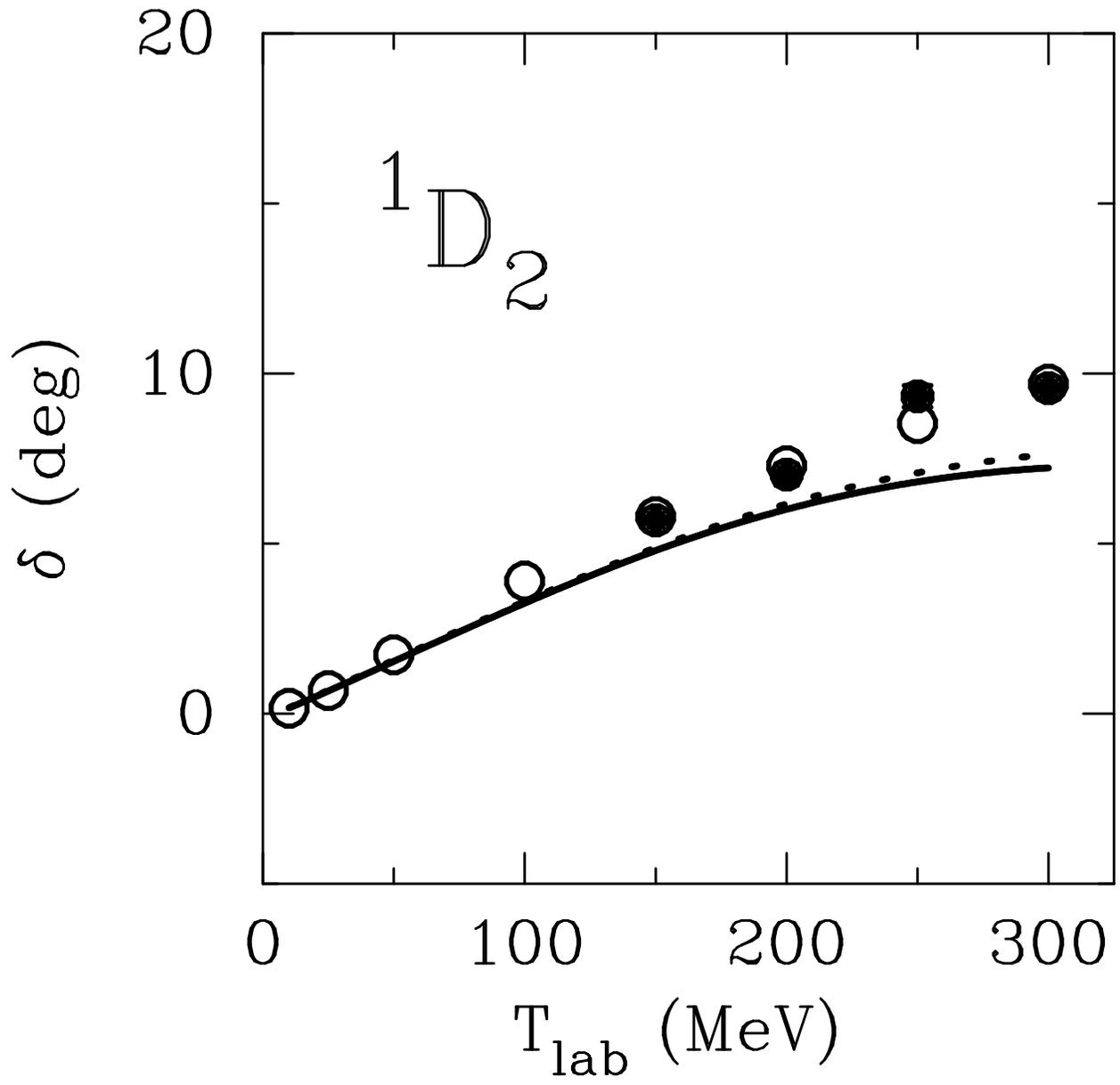
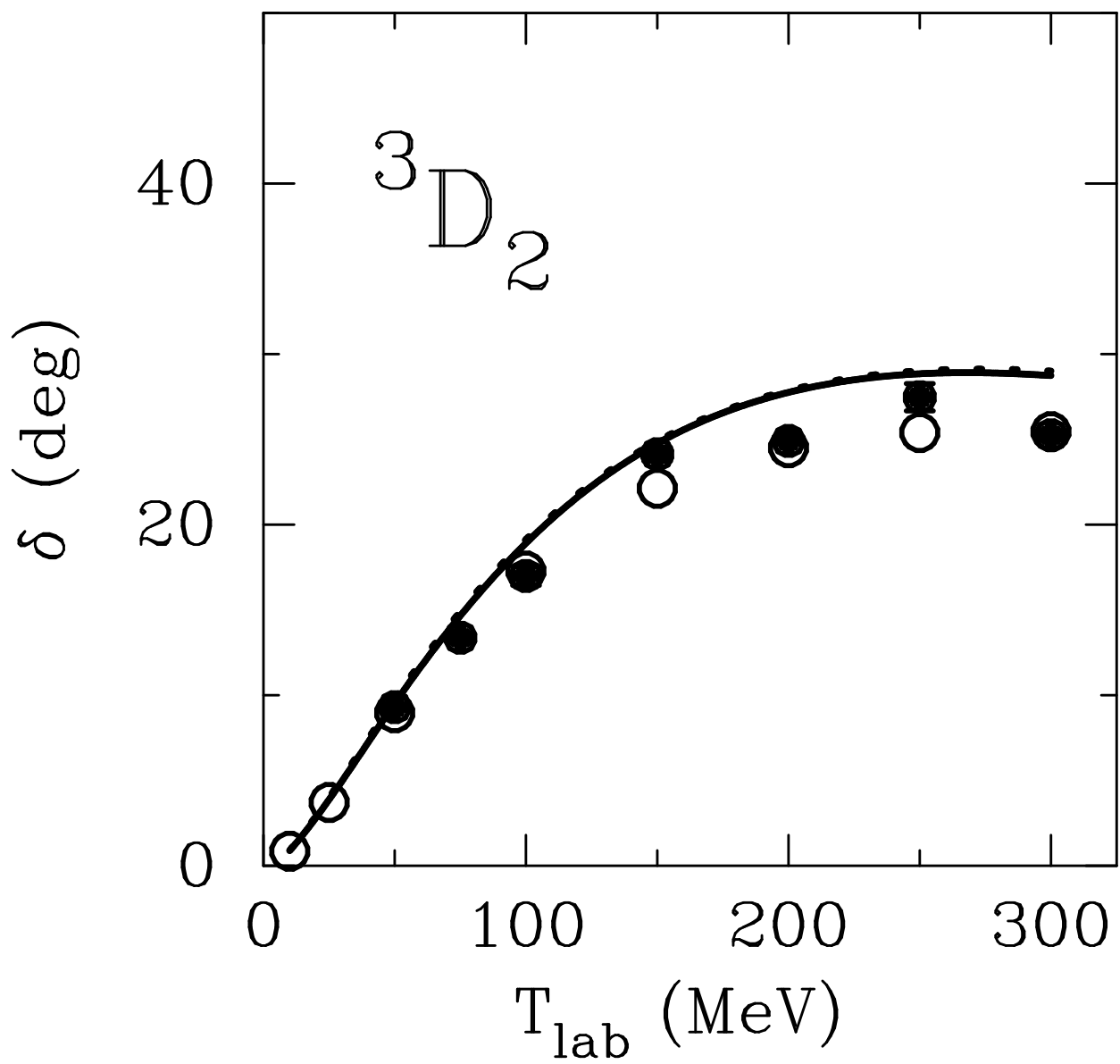
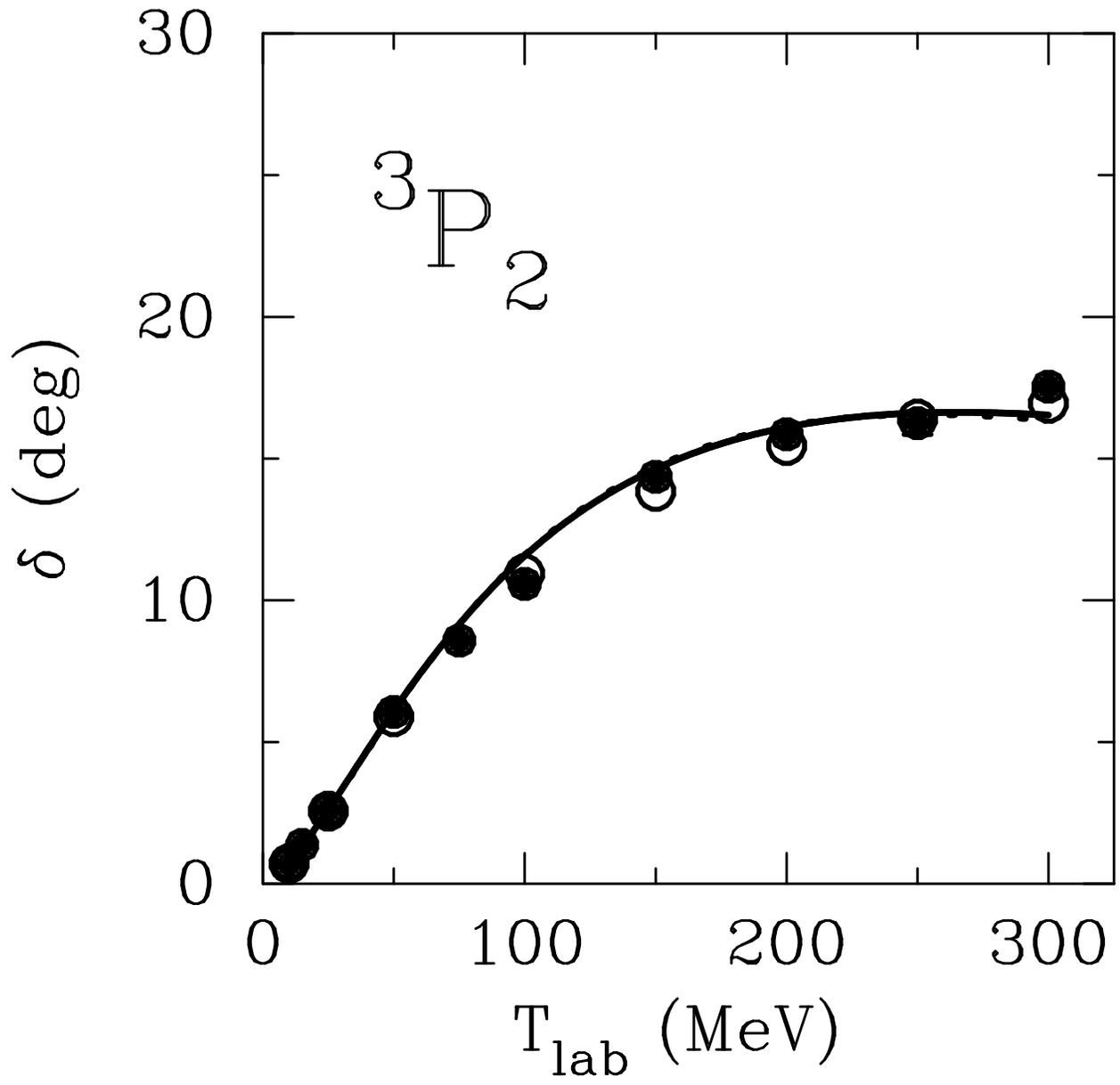
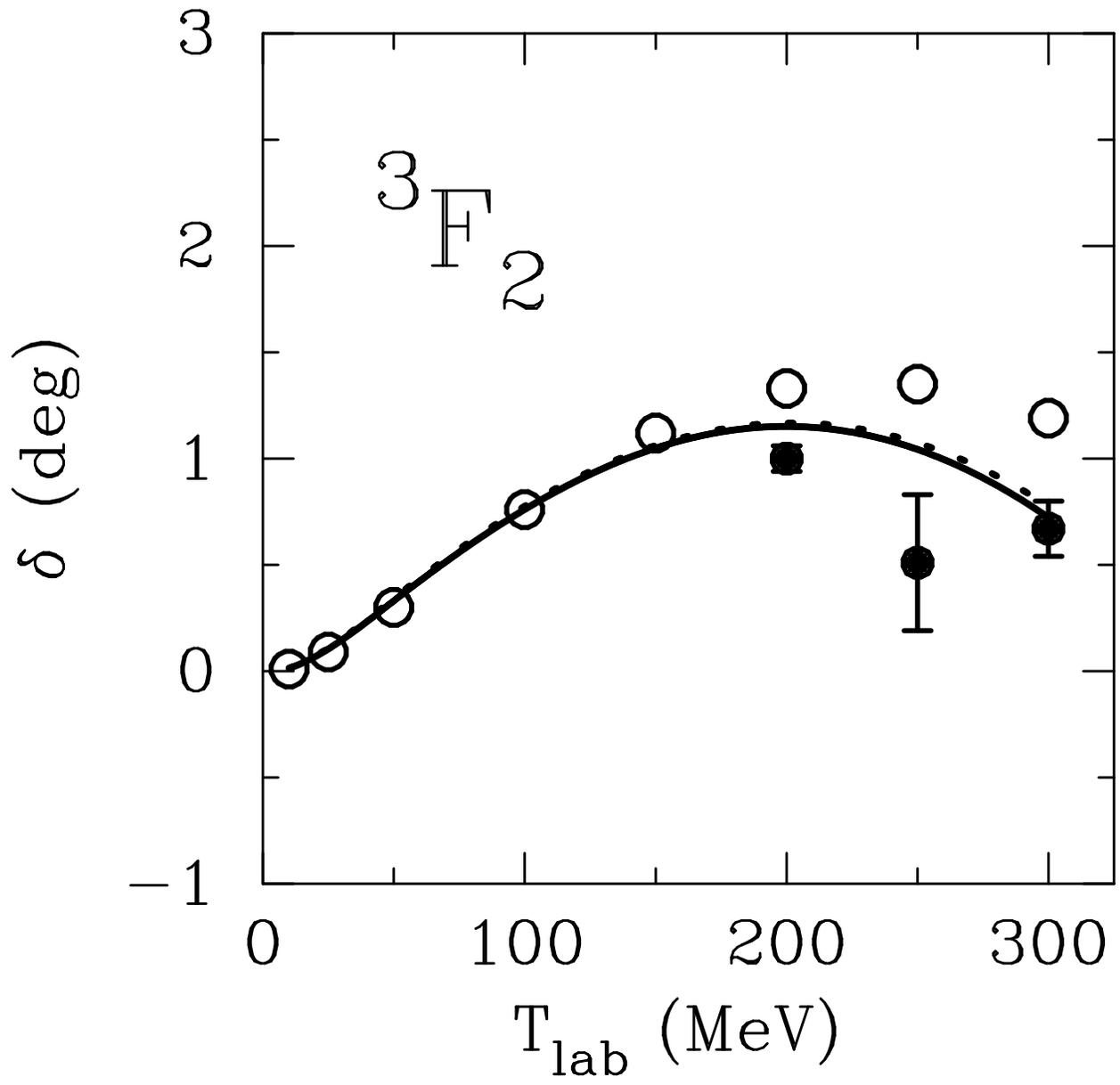
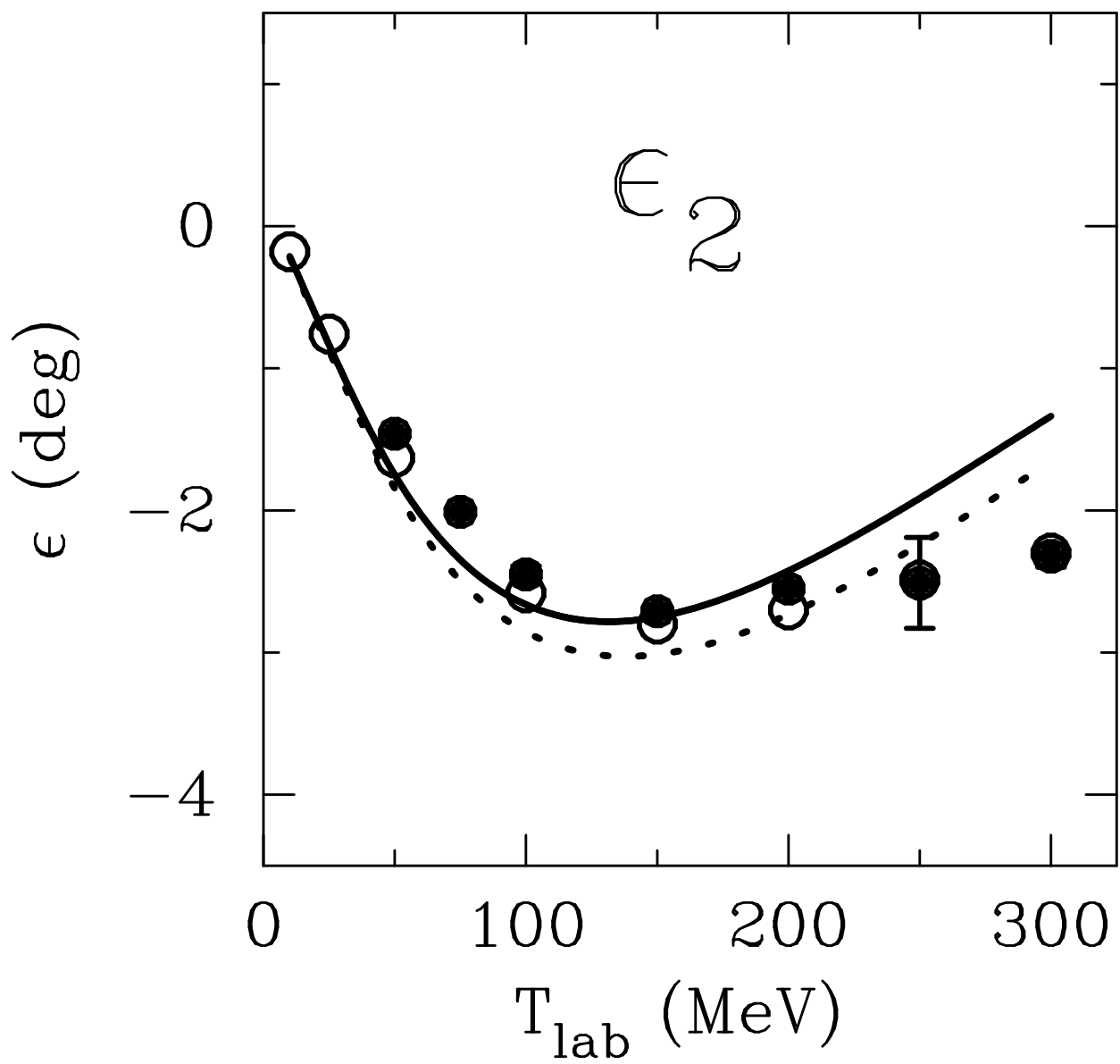
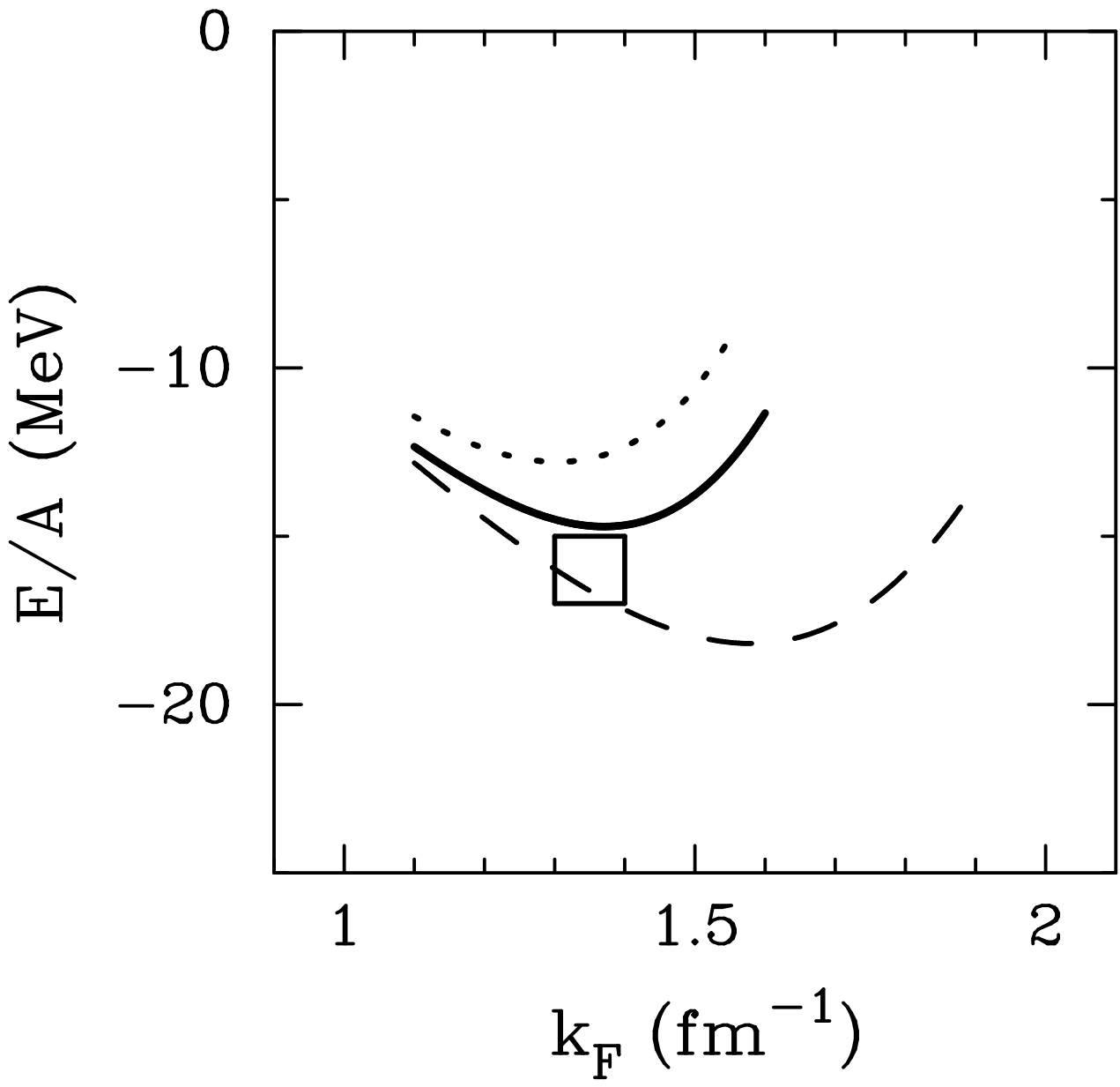
\end{document}